\documentclass[12pt]{article}
\usepackage[latin9]{inputenc}
\usepackage{geometry}
\usepackage{amsmath}
\usepackage{amssymb}
\usepackage{graphicx}
\usepackage{setspace}
\usepackage{esint}
\makeatletter

\usepackage{amsfonts}
\providecommand{\U}[1]{\protect\rule{.1in}{.1in}}

\newtheorem{theorem}{Theorem}\newtheorem{definition}[theorem]{Definition}\newtheorem{lemma}[theorem]{Lemma}\newtheorem{proposition}[theorem]{Proposition}\topmargin0cm
\makeatother

\begin{document}
\title{How To Sell (or Procure) in a Sequential Auction Market\thanks{We thank Paul Klemperer, Tymofiy Mylovanov, Dan Quint, Dan Quigley,
Jószef Sákovics, Vasiliki Skreta, and conference and seminar audiences
for helpful comments and Serafin Grundl and Diwakar Raisingh for excellent
research assistance.}}
\author{Kenneth Hendricks\thanks{Department of Economics, University of Wisconsin, Madison, hendrick@ssc.wisc.edu.}
\and Thomas Wiseman\thanks{Department of Economics, University of Texas at Austin, wiseman@austin.utexas.edu.}}
\date{February 2021}
\maketitle
\begin{abstract}
A seller with one unit of a good faces $N\geq3$ buyers and a single
competitor who sells one other identical unit in a second-price auction
with a reserve price. Buyers who do not get the seller's good will
compete in the competitor's subsequent auction. We characterize the
optimal mechanism for the seller in this setting. The first-order
approach typically fails, so we develop new techniques. The optimal
mechanism features transfers from buyers with the two highest valuations,
allocation to the buyer with the second-highest valuation, and a withholding
rule that depends on the highest two or three valuations. It can be
implemented by a modified third-price auction or a pay-your-bid auction
with a rebate. This optimal withholding rule raises significantly
more revenue than would a standard reserve price. Our analysis also
applies to procurement auctions. Our results have implications for
sequential competition in mechanisms.
\end{abstract}

\section{Introduction}

Sequential auctions are commonly used to sell or procure multiple
products or multiple units of identical products. For example, in
many states, school districts hold procurement auctions for school
milk contracts sequentially during the spring of each year. Auction
houses such as Sotheby's and Christie's sell collections of goods
in a sequence of single-object auctions. Sellers on auction platforms
such as eBay sell their goods individually and these auctions are
sequenced by their unique arrival times. Much of the theoretical literature
on sequential auctions is focused on characterizing the equilibrium
behavior of bidders under the assumption that sellers (or buyers,
in the case of procurement auctions) are passive and nonstrategic.
However, in many of these settings, the products are owned by different
sellers (or demanded by different buyers). In this paper, we analyze
how a seller (or buyer) facing competition in the form of a subsequent
auction can design a mechanism to maximize revenue (or minimize cost).
We will formulate this design problem in terms of sale auctions but
it should be clear from the analysis that it also applies to procurement
auctions.

We consider the design problem in a setting where losing bidders compete
in another, subsequent auction or market, but where the winner has
satisfied his demand and drops out of the competition. Our focus is
on the allocation externality that arises in this setting and the
impact it has on optimal auction design. The revenue that a seller
can collect from buyers is constrained by her need to incentivize
them to participate in her auction. In our setting, these incentive
constraints depend on the bidder's outside option, which is his payoff
from competing in the subsequent market. That payoff is endogeneous:
it depends on the valuations of the other buyers and on which (if
any) of them wins the auction. If the good is not allocated, or not
allocated efficiently, then the losing buyers face stronger competition
in the subsequent market. This creates a negative payoff externality
on bidders by lowering their outside option. Our primary goal in this
paper is to examine how a seller can exploit this externality to extract
more surplus from the bidders.

An important example is the market for pharmaceutical drugs in middle-income
countries such as Ecuador. Brugués \cite{Brugues} develops and estimates
a two-stage model of this market. In the first stage, the Ecuadorian
government runs several hundred, nation-wide procurement auctions
with reserve prices for different drug products.\footnote{Brugués \cite{Brugues} defines products mainly by their main active
ingredient or molecule, and markets based on the Anatomical Therapeutic
Chemical (ATC) classification of drugs. For example, statins is a
market with products of different ingredients, such as atorvastatin
or simvastatin.} In each auction, a small number of firms compete to become the sole
provider of a specific product to every public hospital and pharmacy
in the country for two years. In the second stage, many of these same
firms sell their products to private pharmacies and compete in prices.
Drugs offered in the public sector are free to consumers but involve
significant waiting and prescription renewal costs. Consumers can
acquire these drugs more quickly and easily from private pharmacies,
but they would have to pay out-of-pocket. Despite offering a much
smaller range of products than the private sector, the public sector
accounts for 34\% of total sales in the markets where the government
offers products.

This auction setting features the allocation externality that we study.
Losing bidders compete in the private market, but the winning bidder
is unlikely (and Brugués's \cite{Brugues} results bear this out)
to be a serious competitor in that market because its capacity may
be limited and lower prices reduce demand in the public sector. After
an auction where the reserve price is not met, the government does
not offer that product, and all bidders can compete in the private
sector. The key simplification that we would need to make for our
analysis to apply directly is to assume that firms sell identical
products. Given this assumption, the private market becomes a second-price
auction in which the firm with the lowest cost wins the market at
a price equal to the second-lowest cost. This assumption may not conform
exactly to the actual situation even though in some markets, especially
those with generics, the products may be very close substitutes. Nevertheless,
we believe that our analysis provides some insight into how the government
can redesign its auction to reduce its procurement payments to firms.

Our baseline model is simple. There are two sellers, each with a single
unit of an identical good, who sell their units sequentially to $N\geq3$
buyers. These buyers have unit demands. Their values are private and
independently drawn from a common distribution $F$ with density $f$.
Any buyer who fails to obtain the good from the first seller participates
in the auction of the second seller. The second auction is a second-price
auction. In this case, buyers have a dominant strategy to bid their
value in that auction. Thus, any information that buyers obtain about
their competitors' types from the first auction has no effect on their
bidding behavior in the subsequent auction. The second seller is passive
and non-strategic. He does not adjust the reserve price or the auction
rules in response to the first seller's choice of auction or disclosure
of outcomes. A competitive market for a second unit, as in the Ecuadorean
pharmaceutical example, corresponds to a zero reserve price in the
second auction. So does a setting where, as in Carroll and Segal \cite{CarrollSegal},
types are revealed after the first mechanism runs, and the buyer with
the highest remaining value makes a take-it-or-leave-it offer to the
second seller. Given these assumptions, the first seller's problem
consists of designing an allocation and pricing rule to maximize revenues.

Our first result is a characterization of the optimal direct mechanism.
We are interested in this mechanism because it establishes how much
revenue a seller can achieve and, more importantly, how she can achieve
it. We show that the seller's design problem reduces to a revenue-maximization
problem that can be solved using standard methods from Myerson \cite{Myerson}.
However, the solution is quite different from the optimal mechanism
of a monopoly seller. In the latter case, a seller maximizes revenues
by allocating the good to the buyer with the highest reported value
as long as that report is high enough. By contrast, in our model,
the seller optimally allocates her good only when the \emph{second}-highest
report is large \emph{compared to} the third-highest. This withholding
rule clearly cannot be implemented by a reserve price. The second
difference is that the seller allocates the object (if at all) to
the second-highest bidder rather than the highest. That policy, because
it ensures that the highest-value buyer always participates in the
second auction, eliminates the incentive that a buyer may have to
underreport his value in the hopes of increasing the probability that
the good is allocated to someone else (recall that allocation is more
likely when the third-highest reported value is low) and thus reducing
future competition. Note however that misallocation does not occur.
If the first good is allocated, then the highest two types get the
two goods.

The intuition behind the optimal allocation rule is as follows. Recall
that when there is only one seller, the maximum surplus that she can
create by allocating her good is $x_{(1)},$ the highest value among
the buyers. (The seller's value of the object is zero.) However, because
the values of the buyers are private information, the most that the
seller can extract is $\psi(x_{(1)})$, where $\psi(x)\equiv x-(1-F(x))/f(x)$
is the virtual value of a buyer with value $x$. Thus, the optimal
allocation rule in the single seller case is to allocate only if $\psi(x_{(1)})$
is positive$.$ When there is a second seller, the maximum surplus
that the first seller can extract from the buyers is the difference
between the surpluses generated from allocating the good and from
not allocating it. If the first good is not allocated, then the second
good will go to the buyer with the highest value at a price equal
to the second-highest value, yielding a buyer surplus of $x_{(1)}-x_{(2)}$.
If the highest-value buyer gets the first object, then the second
object will go to the buyer with the second-highest valuation at a
price equal to the third highest, and total surplus for the buyers
is $x_{(1)}+x_{(2)}-x_{(3)}$. (Note that the same total surplus results
if the first object goes to the buyer with the second-highest value,
since then the highest-value buyer gets the second object.) The difference
is $2x_{(2)}-x_{(3)}$. As in the monopoly case, the seller cannot
extract all of that surplus because of the private information of
the buyers. Instead, we show that she can get $\psi(x_{(2)})+x_{(2)}-x_{(3)}.$
The optimal rule, then, is to allocate whenever $\psi(x_{(2)})+x_{(2)}-x_{(3)}$
is positive and not otherwise.

We extend our analysis to the case where the second seller sets a
non-trivial reserve price $r$. A significant complication arises:
the solution to the first seller's optimization problem given first-order
incentive constraints turns out to violate global incentive compatibility
when $r$ is less than $\psi^{-1}(0)$, the optimal reserve price
in the single seller case. Nevertheless, we characterize the solution,
using techniques that, like those of Bergemann et al.\ \cite{BBM}
and Carroll and Segal \cite{CarrollSegal}, may be useful in other
mechanism design settings where the first-order approach fails. We
show that the solution preserves the basic features of the optimal
mechanism for the baseline case. However, the optimal withholding
rule also exhibits interesting new features. It now depends on the
first-, second-, and third-highest values rather than just the second-
and third-highest, and it varies with the number of bidders.

We show that the optimal mechanism can be implemented by a modified
third-price auction or by a pay-your-bid auction in which the highest
bidder gets a rebate equal to the sale price of the second item. In
the third-price auction, it is ex post incentive compatible for buyers
to report their type truthfully; in the pay-your-bid auction, the
equilibrium is in monotone bid functions. Consequently, in each of
these auctions, the seller can use the bids\ to implement the optimal
allocation rule. The unusual feature of these auctions is that both
the highest and second-highest bidder make payments to the seller
when the good is allocated. In the pay-your-bid auction, they simply
pay their bids and, in the third-price auction, their payments are
based on the third-highest bid. The intuition for why the seller can
extract payments from the two highest bidders is that both benefit
from the good being allocated and are willing to pay to ensure that
this event occurs.

We evaluate the revenue gains from using an optimal mechanism in our
baseline case against two benchmarks. One is when the seller must
sell the good with probability one. The expected revenue in this case
is simply the expected value of the third order statistic. (This is
also the expected revenue that the seller can obtain if she uses a
standard first- or second-price auction with no reserve price.) We
use the ``must sell''\ auction as a benchmark for evaluating the
revenue gains from the optimal allocation rule. These gains are substantial.
In our uniform example, we find that the expected revenue to the first
seller increases by 53\% (relative to the third order statistic) when
she uses the optimal mechanism. The second seller also benefits since
the sale price of her good increases from $x_{(3)}$ to $x_{(2)}$
if the first seller does not allocate the good. Her expected revenue
increases by 16\%.

The second benchmark is a standard auction with an optimal reserve
price. We first prove that the presence of the allocation externality
implies that the standard first or second-price auction with a reserve
price does not have a strictly increasing symmetric equilibrium. However,
as Jehiel and Moldovanu \cite{JehielMoldovanu} have shown, there
is an equilibrium with partial pooling at the reserve price. We derive
the partial-pooling equilibrium for our uniform example and compute
the expected revenues from using an optimal reserve price. We find
that a reserve price is not a very effective way for the seller to
raise revenues. The gain in expected revenue is only 21\% (compared
to no reserve price), roughly half of the gain from the optimal auction.
One reason is that the partial pooling is a source of allocative inefficiency,
because it implies that a buyer whose value is below the two highest
may get the good. The other, more important reason is that the outside
option of winning the second auction at a price below the reserve
price causes the participation threshold in the first auction to be
substantially higher than the optimal reserve price. In our example,
only 40\% of the bidders bid in the first auction and roughly half
of them bid the reserve price. Clearly, the threat to withhold the
good if the second-highest bid is too low relative to the third-highest
bid is more effective than the threat to withhold if the highest bid
is too low.

Our paper is the first to study optimal mechanism design in sequential
auctions with competing sellers. Milgrom and Weber \cite{MilgromWeber}
studies sequential auctions with no reserve prices in an IPV environment
with $N$ buyers who have unit demands and shows that prices for $k$
identical objects sold sequentially in first- or second-price forms
a martingale and are on average equal to the expected value of the
$\left(k+1\right)$-th order statistic.\footnote{There is a large empirical literature that tests the martingale prediction
(e.g., Ashenfelter \cite{Ashenfelter}, Ashenfelter and Genesove \cite{AshenfelterGenesove},
and Beggs and Graddy \cite{BeggsGraddy}). Ashenfelter and Graddy
\cite{AshenfelterGraddy} provide a survey of this literature.} Black and de Meza \cite{BlackdeMeza} examine the impact of multi-unit
demands on prices in sequential, second-price auctions in a model
with two passive sellers and two identical goods. Budish and Zeithammer
\cite{BudishZeithammer} use this setting to extend the Milgrom and
Weber analysis to imperfect substitutes (and two-dimensional types).
Kirkegaard and Overgaard \cite{KirkegaardOvergaard} show that the
early seller in the Black and de Meza model can increase her expected
revenue by offering an optimal buy-out price. Our analysis allows
the early seller to consider any mechanism, in the special case of
unit demands.\footnote{There is a growing literature (e.g., Backus and Lewis \cite{BackusLewis},
Said \cite{Said}, and Zeithammer \cite{Zeithammer}) that studies
bidding behavior in sequential, second-price auctions in stationary
environments where new buyers and sellers enter the market each period.
These papers make behavioral assumptions that effectively rule out
the allocation externality.}

This paper is related to the work on auctions with externalities,
where the payoff to a losing bidder depends on whether and to whom
the object is allocated, and to the work on type-dependent outside
options, where bidders have private information about their payoff
if they lose.\footnote{This paper is also related to the recent literature on optimal design
of auctions (and disclosure rules) in which the externalities are
due to resale (e.g., Bergemann et al.\ \cite{BBM}, Calzolari and
Pavan \cite{CalzolariPavan}, Carroll and Segal \cite{CarrollSegal},
Dworczak \cite{Dworczak}, and Virág \cite{Virag}).} Jehiel and Moldovanu \cite{JehielMoldovanu} study the impact of
interactions by buyers in a post-auction market on bidding behavior
in standard auctions. Figueroa and Skreta \cite{FigueroaSkreta} and
Jehiel et al.\ \cite{JehielMoldovanuStacchetti1996,JehielMoldovanuStacchetti1999}
consider revenue-maximizing mechanisms in a more general model of
externalities. In our setting, the payoff to a buyer who fails to
get the first object depends both on his own type and on the highest
value among the other losing buyers. A feature of our environment
is that the optimal threat by the seller -- that is, the action that
minimizes the continuation payoff of all non-participating buyers
-- is to not allocate the object. A consequence is that the participation
constraint binds only for the lowest type of buyer. The optimal threat
in Figueroa and Skreta \cite{FigueroaSkreta} and Jehiel et al.\ \cite{JehielMoldovanuStacchetti1996,JehielMoldovanuStacchetti1999}
is more complicated, and calculating the ``critical type'' for whom
the participation constraint binds can be challenging.

Finally, our paper contributes to the literature on competing mechanisms.
This literature is focused on markets where sellers with identical
goods choose their mechanisms simultaneously and buyers then select
among them. Burguet and Sákovics \cite{BurguetSakovics} study the
case of two sellers with identical goods who simultaneously choose
reserve prices in second-price auctions. They find that competition
for buyers lowers equilibrium reserve prices, but not to zero. McAfee
\cite{McAfee}, Peters and Severinov \cite{PetersSeverinov}, and
Pai \cite{Pai} consider the general mechanism choice problem and
show that, when the number of sellers and buyers in a homogeneous
good market is large, second-price auctions with zero reserve prices
emerge as an equilibrium mechanism. These results lead Peters \cite{Peters2}
to conclude that competition among sellers promotes simple, more efficient
mechanisms. Our results suggest that this conclusion may not apply
when auctions are sequenced. The early seller in our model does not
have to compete for buyers. When he uses the optimal withholding rule,
all buyers participate because, in doing so, they increase the likelihood
that the good is allocated and their chances of winning the subsequent
auction. This is not the case when the seller tries to withhold the
good using a simple reserve price. We discuss competition between
sellers further in Section \ref{sec:Extensions}.

The organization of the rest of the paper is as follows. In Section
\ref{sec:Model} we present the model. In Section \ref{sec:Optimal Mechanism}
we derive the optimal allocation rule when the reserve price in the
second auction is zero, and in the next section we extend the analysis
to the case of a non-trivial reserve price. In Section \ref{sec:Implementation}
we show that the optimal mechanism can be implemented using a modified
third-price auction or a pay-your-bid auction with a rebate. We evaluate
the gains from using the optimal mechanism by comparing it to a standard
auction with and without an optimal reserve price in Section \ref{sec:Revenues}.
In Section \ref{sec:Extensions}, we consider extensions. Section
\ref{sec:Conclusions} provides concluding remarks.

\section{Model\label{sec:Model}}

There are $N$ \emph{ex ante} identical potential buyers, indexed
by $i$, with unit demand for an indivisible good. Each buyer $i$'s
privately observed valuation for the good $X_{i}$ is independently
drawn from distribution $F$ with support $\left[\underline{x},\bar{x}\right]$,
$\underline{x}\geq0$. We will sometimes refer to a buyer's valuation
as his \emph{type}. We assume that $F$ has a continuous density $f$
and that the virtual valuation $\psi(x)\equiv x-(1-F(x))/f(x)$ is
increasing in $x$. Order the valuations from highest to lowest $X_{(1)},X_{(2)},\ldots,X_{(N)}$.

There are two sellers who sell identical units of the good. Each seller
sells one unit. They sell their units sequentially over two periods
and we refer to them in the order that they sell. The second seller
uses a second-price auction with reserve price $r\geq0$. Given $r$,
the first seller chooses his mechanism. Both sellers' valuations of
the good are normalized to zero. This structure is common knowledge.
We will characterize the revenue-maximizing mechanism for the seller
in the first period, given that any buyer who does not obtain the
first object will participate in the auction for the second object.
In what follows, we typically refer to the first seller as just ``the
seller.\textquotedblright{}

In our model, it is a weakly dominant strategy for any buyer who did
not obtain the first object to submit a bid equal to his valuation
in the second auction. Thus, in designing his mechanism, the seller
does not have to be concerned about the leakage problem. Any information
buyers acquire in the first period about the types of competitors
does not influence their bidding behavior in the second period. As
a result, buyers have no incentive to bid untruthfully in period one
to affect behavior in period two. However, in period one, a buyer's
bid may still influence the allocation of the first object, which
does affect outcomes in the second period. The design of the revenue-maximizing
mechanism for the seller must take that incentive into account.

Without loss of generality, we restrict attention to direct mechanisms
in which buyers report their types. Let $\mathbf{x}\in\left[\underline{x},\bar{x}\right]^{N}$
denote the vector of \textit{reported} types. A direct mechanism in
our context specifies, for any given $\mathbf{x}$, the probability
that each bidder $i$ gets the good is $P_{i}(\mathbf{x)\geq0}$ with
$\Sigma_{i=1}^{N}P_{i}(\mathbf{x)\leq1}$ and the payment $t_{i}(\mathbf{x)}$
that he must make.

We will work quite a bit with order statistics. For $k\in\left\{ 1,\ldots,N\right\} $,
let $F_{k}(x)$ denote the distribution of the $k$-th order statistic
$X_{(k)}$, and let $f_{k}(x)$ denote the corresponding density.
We will also need to define the distribution of an order statistic
conditional of the value of another order statistic. Let $F_{k|x_{(j)}}$
and $f_{k|x_{(j)}}$ denote the distribution and density, respectively,
of the $k$-th order statistic conditional on the value of the $j$-th
order statistic $X_{(j)}=x_{(j)}$ for $j\neq k$.

Finally, it will also be useful to define the order statistics of
the competing valuations that a single buyer faces. Order the valuations
of the other $N-1$ buyers from highest to lowest $Y_{(1)},Y_{(2)},\ldots,Y_{(N-1)}$.
We denote the distributions of $Y_{(k)}$ by $G_{k}(x)$, and the
corresponding density by $g_{k}(x)$. The conditional distributions
and densities of order statistics among a bidder's rivals, $G_{k|y_{(j)}}$
and $g_{k|y_{(j)}}$ for $j\neq k$, are defined analogously.

\section{The Optimal Mechanism when $r=0$\label{sec:Optimal Mechanism}}

We begin by assuming no reserve price in the second auction (or, equivalently,
that $r\leq\underline{x}$). The payoff to a buyer $i$ with valuation
$X_{i}$ in the second period, provided that he did not obtain the
first object, depends on whether or not the first object was allocated
to the competitor with the highest type $Y_{(1)}$. If so, then buyer
$i$'s payoff, $\max\left\{ X_{i}-Y_{(2)},0\right\} $, is a function
of the highest remaining competitor's type $Y_{(2)}$. If not, then
buyer $i$'s payoff is $\max\left\{ X_{i}-Y_{(1)},0\right\} $. All
else equal, buyer $i$ prefers that the first object go to his strongest
competitor so that competition in the subsequent auction is reduced.
Thus, the expected payoff to a buyer depends on the two highest valuations
among his competitors. We denote the highest-type competitor of bidder
$i$ by $j(1)$ (so that $X_{j(1)}=Y_{(1)}$). Then, the expected
payoff to a bidder $i$ with type $x_{i}$ given vector of reports
$\mathbf{x}$, excluding any payment to the first seller, is 
\begin{equation}
P_{i}(\mathbf{x)}\cdot x_{i}+P_{j(1)}(\mathbf{x)}\cdot\max\left\{ x_{i}-y_{(2)},0\right\} +\left(1-P_{i}(\mathbf{x)}-P_{j(1)}(\mathbf{x)}\right)\cdot\max\left\{ x_{i}-y_{(1)},0\right\} .\label{eq:NewPayoff}
\end{equation}
To interpret Expression \ref{eq:NewPayoff}, observe that if $x_{i}$
is not one of the two highest valuations (if $y_{(2)}>x_{i}$), then
bidder $i$ gets a payoff only if he receives the first object. If
bidder $i$ has the second-highest valuation (if $y_{(1)}>x_{i}>y_{(2)}$),
then he again receives his valuation if the first object is allocated
to him, but he also gets payoff $x_{i}-y_{(2)}$ from winning the
second auction if the first object goes to bidder $j(1)$. Finally,
if $x_{i}$ is the highest valuation (if $x_{i}>y_{(1)})$, then bidder
$i$ either 1) gets the first object, 2) gets the second object at
price $y_{(2)}$ if the first object goes to bidder $j(1)$, or 3)
gets the second object at price $y_{(1)}$ in any other case.

Next, we use the first-order incentive compatibility constraints to
express the transfer payments from buyers in terms of their payoffs
and the allocation rule, and then choose the allocation rule to maximize
the sum of payments. The standard approach defines the payoffs and
allocation rule in terms of the vector of reported types. However,
in our case, a bidder's payoff depends not only upon reported types
but also upon the highest actual types among his competitors who do
not get the first object. This dependence creates problems summing
Expression \ref{eq:NewPayoff} across bidders because the set of competitors
varies with the identity of the bidder. To deal with this issue, we
exploit the symmetry of the bidders and re-define payoffs and allocations
in terms of the vector of reported realizations of order statistics.
For any vector of reported types $\mathbf{x}$, define $\hat{\mathbf{x}}$
as the vector of reported types ordered from highest to lowest (with
ties broken arbitrarily). Thus, the $k$-th element of $\hat{\mathbf{x}}$
is the $k$-th highest reported type in $\mathbf{x}$ (i.e., $\widehat{x}_{k}=x_{(k)})$.
Let $\hat{\mathbf{f}}$ denote the joint density of $\widehat{\mathbf{x}}$.

Similarly, let $\widehat{\mathbf{y}}$ denote the ordered vector of
competitors' reported types facing a single buyer, with joint density
$\widehat{\mathbf{g}}$. Given a bidder's type $x$ and competitors'
types $\widehat{\mathbf{y}}$, let $\left(x;\widehat{\mathbf{y}}\right)$
denote the ordered vector of all $N$ types.

We begin with the allocation rule. For each $k\in\{1,\ldots,N\}$,
let $\hat{p}^{k}(\hat{\mathbf{x}})$ denote the probability that the
mechanism allocates the object to the bidder with the $k$-th highest
report, given $\mathbf{\hat{\mathbf{x}}}$. Assuming that other buyers
report truthfully, we can then write the interim expected payoff to
a buyer of type $x$ who reports truthfully as follows:\footnote{For completeness, set $\widehat{y}_{k+1}=\underline{v}$ when $k=N-1$.}
\begin{equation}
\begin{array}{ccc}
\Pi(x|x) & = & {\int\limits _{\left[\underline{x},\bar{x}\right]^{N-1}:x>\widehat{y}_{1}}}\left(x-\widehat{y}_{1}+\hat{p}^{1}\left(\left(x;\widehat{\mathbf{y}}\right)\right)\cdot\widehat{y}_{1}+\hat{p}^{2}\left(\left(x;\widehat{\mathbf{y}}\right)\right)\cdot\left[\widehat{y}_{1}-\widehat{y}_{2}\right]\right)\widehat{\mathbf{g}}(\widehat{\mathbf{y}})\\
 & + & {\int\limits _{\left[\underline{x},\bar{x}\right]^{N-1}:\widehat{y}_{1}\geq x>\widehat{y}_{2}}}\left(\hat{p}^{1}\left(\left(x;\widehat{\mathbf{y}}\right)\right)\cdot\left[x-\widehat{y}_{2}\right]+\hat{p}^{2}\left(\left(x;\widehat{\mathbf{y}}\right)\right)\cdot x\right)\widehat{\mathbf{g}}(\widehat{\mathbf{y}})\\
 & + & {\displaystyle \sum_{k=2}^{N-1}}\left[{\int\limits _{\left[\underline{x},\bar{x}\right]^{N-1}:\widehat{y}_{k}\geq x>\widehat{y}_{k+1}}}\left(\hat{p}^{k+1}\left(\left(x;\widehat{\mathbf{y}}\right)\right)\cdot x\right)\widehat{\mathbf{g}}(\widehat{\mathbf{y}})\right].
\end{array}\label{expected gross payoff}
\end{equation}
More generally, in the appendix we derive the payoff $\Pi(q|x)$ to
a buyer of type $x$ who reports his type as $q$. We further show
that $\Pi(q|x)$ is convex in the valuation; that is, $\Pi_{22}(q|x)\geq0$.

The next steps are standard. Let $t(q)$ be the expected transfer
to the seller from a buyer who reports type $q$. Incentive compatibility
requires that buyers report their valuations truthfully, so the equilibrium
payoff to a buyer of type $x$ is 
\[
U(x)=\max_{q}\Pi(q|x)-t(q).
\]
As the maximum of convex functions, $U(x)$ also is convex. It is
therefore absolutely continuous and so differentiable almost everywhere.
By standard arguments, its derivative is given by $U^{\prime}(x)=\Pi_{2}(x|x)$,
and 
\begin{equation}
U(x)=U(\underline{x})+
{\textstyle \int\limits _{\underline{x}}^{x}}
\Pi_{2}(x^{\prime}|x^{\prime})dx^{\prime},\label{eq:IC rewritten}
\end{equation}
where $\Pi_{2}(x|x)$ is the partial derivative of $\Pi(q|x)$ with
respect to the second argument (the buyer's true type) evaluated at
the truthful report. It is given by
\begin{equation}
\begin{array}{ccc}
\Pi_{2}(x|x) & = & G_{1}(x)+{\int\limits _{\left[\underline{x},\bar{x}\right]^{N-1}:\widehat{y}_{1}\geq x>\widehat{y}_{2}}}\left(\hat{p}^{1}\left(\left(x;\widehat{\mathbf{y}}\right)\right)+\hat{p}^{2}\left(\left(x;\widehat{\mathbf{y}}\right)\right)\right)\widehat{\mathbf{g}}(\widehat{\mathbf{y}})\\
 & + & {\displaystyle \sum_{k=2}^{N-1}}\left[{\int\limits _{\left[\underline{x},\bar{x}\right]^{N-1}:\widehat{y}_{k}\geq x>\widehat{y}_{k+1}}}\left(\hat{p}^{k+1}\left(\left(x;\widehat{\mathbf{y}}\right)\right)\right)\widehat{\mathbf{g}}(\widehat{\mathbf{y}})\right].
\end{array}\label{eq:Pi2}
\end{equation}
Substituting $U(x)=\Pi(x|x)-t(x)$ into Expression \ref{eq:IC rewritten}
then yields 
\begin{equation}
t(x)=t(\underline{x})+\Pi(x|x)-\Pi(\underline{x}|\underline{x})-
{\textstyle \int\limits _{\underline{x}}^{x}}
\Pi_{2}(x^{\prime}|x^{\prime})dx^{\prime}.\label{eq:transfer}
\end{equation}

The mechanism is incentive compatible if for any type $x$ and any
reports $q,q'$ such that $q>x>q'$, we have $\Pi_{2}(q|x)\geq\Pi_{2}(x|x)\geq\Pi_{2}(q'|x)$.
Because allocating the first object to any buyer weakly increases
the total payoff to every buyer (ignoring any period-one transfer),
withholding the first object minimizes the buyers' payoffs. Thus,
the period-one individual rationality condition is that $U(x)$ exceeds
the expected payoff that a buyer of type $x$ could get from the second
auction given that the first seller allocates his unit to no one.
As usual, incentive compatibility implies that the mechanism is individually
rational for all types if it is individually rational for a buyer
of the lowest type $\underline{x}$: $t(\underline{x})\leq\Pi(\underline{x}|\underline{x})$.

The seller's expected revenue is $N\cdot Et(X)$, where the \emph{ex
ante} expected transfer from a buyer is 
\begin{align*}
Et(X) & =t(\underline{x})-\Pi(\underline{x}|\underline{x})+
{\textstyle \int\limits _{\underline{x}}^{\bar{x}}}
\Pi(x|x)f(x)-
{\textstyle \int\limits _{\underline{x}}^{\bar{x}}}
{\textstyle \int\limits _{\underline{x}}^{x}}
\Pi_{2}(x^{\prime}|x^{\prime})dx^{\prime}f(x)dx\\
 & =t(\underline{x})-\Pi(\underline{x}|\underline{x})+
{\textstyle \int\limits _{\underline{x}}^{\bar{x}}}
\left[\Pi(x|x)-\frac{(1-F(x))}{f(x)}\Pi_{2}(x|x)\right]f(x)dx.
\end{align*}
The first equality comes from using Expression \ref{eq:transfer}
and the second from changing the order of integration of the double
integral term. Substituting Expression \ref{expected gross payoff}
for $\Pi(x|x)$ and Expression \ref{eq:Pi2} for $\Pi_{2}(x|x),$
the expected transfer can be expressed in terms of virtual valuations
as follows:
\begin{align}
\begin{array}{c}
Et(X)=t(\underline{x})-\Pi(\underline{x}|\underline{x})\\
+
{\displaystyle \int\limits _{\underline{x}}^{\overline{x}}}
\left\{ \begin{array}{cc}
 & {\int\limits _{\left[\underline{x},\bar{x}\right]^{N-1}:x>\widehat{y}_{1}}}\left(\psi(x)-\widehat{y}_{1}+\hat{p}^{1}\left(\left(x;\widehat{\mathbf{y}}\right)\right)\cdot\widehat{y}_{1}+\hat{p}^{2}\left(\left(x;\widehat{\mathbf{y}}\right)\right)\cdot\left[\widehat{y}_{1}-\widehat{y}_{2}\right]\right)\widehat{\mathbf{g}}(\widehat{\mathbf{y}})\\
+ & {\int\limits _{\left[\underline{x},\bar{x}\right]^{N-1}:\widehat{y}_{1}\geq x>\widehat{y}_{2}}}\left(\hat{p}^{1}\left(\left(x;\widehat{\mathbf{y}}\right)\right)\cdot\left[\psi(x)-\widehat{y}_{2}\right]+\hat{p}^{2}\left(\left(x;\widehat{\mathbf{y}}\right)\right)\cdot\psi(x)\right)\widehat{\mathbf{g}}(\widehat{\mathbf{y}})\\
+ & {\displaystyle \sum_{k=2}^{N-1}}\left[{\int\limits _{\left[\underline{x},\bar{x}\right]^{N-1}:\widehat{y}_{k}\geq x>\widehat{y}_{k+1}}}\left(\hat{p}^{k+1}\left(\left(x;\widehat{\mathbf{y}}\right)\right)\cdot\psi(x)\right)\widehat{\mathbf{g}}(\widehat{\mathbf{y}})\right]
\end{array}\right\} f(x).
\end{array}\label{eq.rewritten revenue}
\end{align}

Note that the probability that a given bidder has the $k$-th highest
value is $1/N$ for each $k\in\left\{ 1,\ldots,N\right\} $. Therefore,
we can rewrite the expected transfer from each bidder in Expression
\ref{eq.rewritten revenue} as 
\begin{align}
Et(\widehat{X}) & =t(\underline{x})-\Pi(\underline{x}|\underline{x})\label{eqrevenue}\\
 & +\frac{1}{N}\int_{\left[\underline{x},\bar{x}\right]^{N}}\left(\psi(\widehat{x}_{1})-\widehat{x}_{2}+\hat{p}^{1}(\widehat{\mathbf{x}})\cdot\widehat{x}_{2}+\hat{p}^{2}(\widehat{\mathbf{x}})\cdot\left[\widehat{x}_{2}-\widehat{x}_{3}\right]\right)\mathbf{\hat{\mathbf{f}}}\left(\widehat{\mathbf{x}}\right)\nonumber \\
 & +\frac{1}{N}\int_{\left[\underline{x},\bar{x}\right]^{N}}\left(\hat{p}^{1}(\widehat{\mathbf{x}})\cdot\left[\psi\left(\widehat{x}_{2}\right)-\widehat{x}_{3}\right]+\hat{p}^{2}(\widehat{\mathbf{x}})\cdot\psi\left(\widehat{x}_{2}\right)\right)\mathbf{\hat{\mathbf{f}}}\left(\widehat{\mathbf{x}}\right)\nonumber \\
 & +\frac{1}{N}\sum_{k=3}^{N}\int_{\left[\underline{x},\bar{x}\right]^{N}}\left(\hat{p}^{k}(\widehat{\mathbf{x}})\cdot\psi\left(\widehat{x}_{k}\right)\right)\mathbf{\hat{\mathbf{f}}}\left(\widehat{\mathbf{x}}\right).\nonumber 
\end{align}
The seller maximizes expected revenue $ER(\widehat{X})=N\cdot Et(\widehat{X})$
subject to incentive compatibility and individual rationality. To
find the optimal allocation rule, we ignore the constraints and maximize
the integral pointwise. Given any vector of ordered types $\hat{\mathbf{x}}$,
taking the derivative of the seller's expected revenue with respect
to $\hat{p}^{k}\left(\hat{\mathbf{x}}\right)$ yields 
\begin{align*}
\frac{\partial ER(\widehat{X})}{\delta\hat{p}^{1}\left(\hat{\mathbf{x}}\right)} & =\frac{\partial ER(\widehat{X})}{\delta\hat{p}^{2}\left(\hat{\mathbf{x}}\right)}=\left[\psi\left(\widehat{x}_{2}\right)+\widehat{x}_{2}-\widehat{x}_{3}\right]\mathbf{\hat{\mathbf{f}}}(\widehat{\mathbf{x}}),
\end{align*}
and for all $k>2$, 
\begin{align*}
\frac{\partial ER(\widehat{X})}{\delta\hat{p}^{k}\left(\hat{\mathbf{x}}\right)} & =\psi\left(\widehat{x}_{k}\right)\mathbf{\hat{\mathbf{f}}}(\widehat{\mathbf{x}}).
\end{align*}

There are two things to note about the derivatives. First, the marginal
revenue from increasing the probability $\hat{p}^{1}$ of allocating
to the highest bidder is exactly the same as from increasing the probability
$\hat{p}^{2}$ of allocating to the second highest bidder. Both are
$\psi(x_{(2)})+x_{(2)}-x_{(3)}$. Intuitively, allocating the unit
to either bidder means that both will obtain a good since the other
bidder gets the second good at the third-highest valuation, $x_{(3)}$.
Not allocating the good means that only the highest bidder will get
a good (the second one), and he will pay the second-highest valuation,
$x_{(2)}$. The difference in surplus between the first case ($x_{(1)}+x_{(2)}-x_{(3)}$)
and the second case ($x_{(1)}-x_{(2)}$) is $2x_{(2)}-x_{(3)}$. Leaving
some surplus for the buyers to incentivize truth-telling results in
replacing one of the $x_{(2)}$ terms with the corresponding virtual
valuation $\psi(x_{(2)})$, and so the seller's marginal revenue is
$\psi(x_{(2)})+x_{(2)}-x_{(3)}$.

The second thing to note is that the marginal benefit from increasing
$\hat{p}^{1}$ or $\hat{p}^{2}$ exceeds the marginal benefit from
increasing the probability $\hat{p}^{k}$ of allocating to any lower-ranked
bidder $k>2$. Because $x_{(2)}\geq x_{(3)}$ and the virtual valuation
$\psi(\cdot)$ is increasing, we have 
\[
\psi(x_{(2)})+x_{(2)}-x_{(3)}\geq\psi(x_{(k)}),
\]
with strict inequality if $x_{(2)}>x_{(k)}$. The solution to the
seller's maximization problem, then, is to allocate to one of the
top two bidders as long as 
\begin{equation}
\psi(x_{(2)})+x_{(2)}-x_{(3)}\geq0,\label{eq:Reserve Rule}
\end{equation}
and not to allocate otherwise. That is, the reserve rule is a function
of the second- and third-highest valuations. The unit is allocated
for certain if $\psi(x_{(2)})\geq0$ because in that case the inequality
holds. If $x_{(2)}+\psi(x_{(2)})<0$, then the unit is certain not
to be allocated. Otherwise, it may or may not be allocated, depending
on the realization of the third order statistic. To maximize revenue,
then, the seller should set $\hat{p}^{1}+\hat{p}^{2}=1$ whenever
$\psi(x_{(2)})+x_{(2)}-x_{(3)}\geq0$, and should set $\hat{p}^{k}=0$
for all $k$ otherwise.\footnote{If $\psi(\underline{x})$, the virtual valuation of the lowest possible
type, is positive, then the object is always allocated, because $\psi(x_{(2)})+x_{(2)}\geq x_{(2)}\geq x_{(3)}$.
In this case, allocating to either the highest or second-highest bidder
is incentive compatible.}

We need to check that the solution to the relaxed problem satisfies
the constraints. The above argument implies that $t(\underline{x})=\Pi(\underline{x}|\underline{x})=0$,
so individual rationality for a buyer with the lowest possible valuation
is satisfied. For incentive compatibility, we want to show that $\Pi_{2}(q|x)\geq\Pi_{2}(x|x)$
for $q>x$ and that $\Pi_{2}(q'|x)\leq\Pi_{2}(x|x)$ for $q'<x$.
It turns out that there is a subtlety relative to the standard mechanism
design environment. Those conditions correspond to the requirement
that that a bidder cannot increase the total probability that he wins
a unit, either the first or the second object, by underreporting his
type, or decrease the probability by overreporting his type. Allocating
to the second-highest bidder (conditional on the good being allocated
at all) satisfies that requirement, but allocating to the top bidder
may not.

The reason that assigning the object to the bidder with the highest
valuation may violate incentive compatibility comes from the fact
that the condition for allocating the good, Expression \ref{eq:Reserve Rule},
is decreasing in the third-highest report $x_{(3)}$. Consider bidder
$i$ with valuation $x_{i}$. Reporting $x^{\prime}<x_{i}$ can raise
the probability that the first good is allocated if $x^{\prime}$
is the third-lowest report. If the unit is assigned to the second-highest
bidder, then allocating it does \emph{not} help bidder $i$ in the
second auction -- assigning it does nothing to reduce competition
in the second auction, because the highest valuation among the remaining
bidders is unchanged. On the other hand, allocating the first unit
to the highest bidder \emph{would} reduce competition in the second
auction. Thus, assigning the unit to the highest bidder can create
a situation where, when $x_{i}$ is the second-highest value, bidder
$i$ would gain from misreporting: if $x_{i}+\psi(x_{i})<x_{(3)}$
but $x_{(3)}+\psi\left(x_{(3)}\right)>x^{\prime}$. Reporting truthfully
means that bidder $i$ does not get a good (the first unit will not
be allocated and the highest bidder will get the second), but by reporting
$x^{\prime}$ bidder $i$ gets the second good (after the first good
is allocated to the highest bidder).

Thus, allocating to the second-highest bidder rather than the first
when Expression \ref{eq:Reserve Rule} is satisfied ensures that the
mechanism is incentive compatible. (Details are in Appendix \ref{sec:Convexity-and-Incentive}.)
Theorem \ref{Theorem: optimal mechanism} summarizes the optimal mechanism.
To describe the transfers concisely, we introduce the following notation.

\begin{definition} For $x\in\left[\underline{x},\bar{x}\right]$,
define $a(x)\in\left[\underline{x},\bar{x}\right]$ as $a(x)\equiv\min\left\{ a\geq x:a+\psi(a)\geq x\right\} .$

\end{definition}That is, $a\left(x_{(3)}\right)$ is the smallest
value of $x_{(2)}$ such that $\psi(x_{(2)})+x_{(2)}-x_{(3)}\geq0$.
Note that $a(x)>x$ when $\psi\left(x\right)<0$ and $a(x)=x$ when
$\psi\left(x\right)\geq0$.

\begin{theorem} \label{Theorem: optimal mechanism}If the distribution
of buyer values $F$ has increasing virtual values and there is no
reserve price in the second auction, then the following is an optimal
(direct) mechanism for the first seller. (Ties are broken randomly.)
\begin{enumerate}
\item \textbf{\textit{Allocation rule:}}\textit{\ The seller allocates
the good to the bidder with the second-highest valuation if} $\psi(x_{(2)})+x_{(2)}-x_{(3)}\geq0$\textit{,
and does not allocate otherwise.}
\item \textbf{\textit{Transfers:}}
\begin{enumerate}
\item \textit{If $\psi\left(x_{(3)}\right)<0$ and the good is allocated
($x_{(2)}\geq a\left(x_{(3)}\right)$), then the bidder with the highest
valuation pays $a\left(x_{(3)}\right)-x_{(3)}>0$, the bidder with
the second-highest valuation pays $a\left(x_{(3)}\right)>0$, and
the other bidders pay nothing.}
\item \textit{If $\psi\left(x_{(3)}\right)\geq0$ (in which case the good
is allocated because $x_{(2)}\geq a\left(x_{(3)}\right)=x_{(3)}$),
then the bidder with the second-highest valuation pays $x_{(3)}>0$
and the other bidders pay nothing.}
\item \textit{If the good is not allocated, then there are no payments.}
\end{enumerate}
\item \textbf{\textit{Revenue:}}\textit{\ The expected revenue to the seller
is} $E\max\left\{ \psi(X_{(2)})+X_{(2)}-X_{(3)},0\right\} .$
\end{enumerate}
\end{theorem}

The transfers come from plugging the allocation rule into Expression
\ref{eq:transfer}. Paralleling the standard mechanism design setting,
a bidder who gets an object (either the first or the second) pays
a transfer equal to his gross payoff minus the gap between his valuation
and the smallest valuation at which he would still get an item, holding
fixed the types of the other bidders. For example, suppose that \textit{$\psi\left(x_{(3)}\right)<0$}\textit{\emph{
and $x_{(2)}\geq a\left(x_{(3)}\right)$, so that the first object
is allocated. Then the bidder with the highest type wins the second
auction and gets a gross payoff of $x_{(1)}-x_{(3)}$. The lowest
valuation at which he would get an object is $a\left(x_{(3)}\right)$:
if his valuation were below $a\left(x_{(3)}\right)$, then the first
object would not be allocated and he would lose the second auction.
Thus, his payment is $x_{(1)}-x_{(3)}-\left[x_{(1)}-a\left(x_{(3)}\right)\right]=a\left(x_{(3)}\right)-x_{(3)}$.}}

The expected revenue expression is obtained by substituting the optimal
allocation rule into Expression \ref{eqrevenue}, integrating, and
recognizing that $E[\psi(X_{(1)}]=E[X_{(2)}].$ We can also write
it in integral form. The allocation rule specifies that the object
is always allocated when $x_{(2)}\geq\psi^{-1}(0)$ and never allocated
when $x_{(2)}<a(\underline{x})<\psi^{-1}(0)$. Then expected revenue
is
\begin{align}
 & {\int_{a(\underline{x})}^{\psi^{-1}(0)}\int_{\underline{x}}^{x_{(2)}+\psi(x_{(2)})}\left[\psi(x_{(2)})+x_{(2)}-x_{(3)}\right]f_{3|x_{(2)}}\left(x_{(3)}\right)f_{2}(x_{(2)})dx_{(3)}dx_{(2)}}\label{expectedrevenue}\\
 & +{\int_{\psi^{-1}(0)}^{\bar{x}}\int_{\underline{x}}^{x_{(2)}}\left[\psi(x_{(2)})+x_{(2)}-x_{(3)}\right]f_{3|x_{(2)}}\left(x_{(3)}\right)f_{2}(x_{(2)})dx_{(3)}dx_{(2)}.}\nonumber 
\end{align}

An interesting benchmark for evaluating the revenue gains from using
the optimal reserve rule is the expected revenue that the seller can
obtain when he must sell the unit with probability 1. It follows from
the above analysis that $\hat{p}^{2}=1$ in the optimal ``must sell\textquotedblright \ mechanism
and that the expected revenue of this mechanism is equal to 
\begin{equation}
E\left[\psi(X_{(2)})+X_{(2)}-X_{(3)}\right].\label{mustsellrevenue}
\end{equation}
The next lemma, which follows from Loertscher and Marx's \cite{LoertscherMarx}
Lemma 1, allows us to express that revenue in terms of expected values
of order statistics.

\begin{lemma} $E[\psi\left(X_{(2)}\right)]=2E[X_{(3)}]-E[X_{(2)}].$
\end{lemma}

\noindent Applying this lemma to Expression \ref{mustsellrevenue}
yields $E[X_{(3)}].$ This result is not too surprising. In our setting,
Milgrom and Weber \cite{MilgromWeber} show that the expected revenue
to the seller who uses a first-price or second-price auction with
no reserve is $E[X_{(3)}].$ The optimal ``must sell\textquotedblright \ mechanism
is revenue equivalent to a first or second-price auction with no reserve
price.

\subsection{Example: Three bidders, uniform valuations\label{sub:Example:-Three-bidders,}}

\begin{figure}[ptb]
\noindent \centering{}\includegraphics[scale=0.4,bb = 0 0 200 100, draft, type=eps]{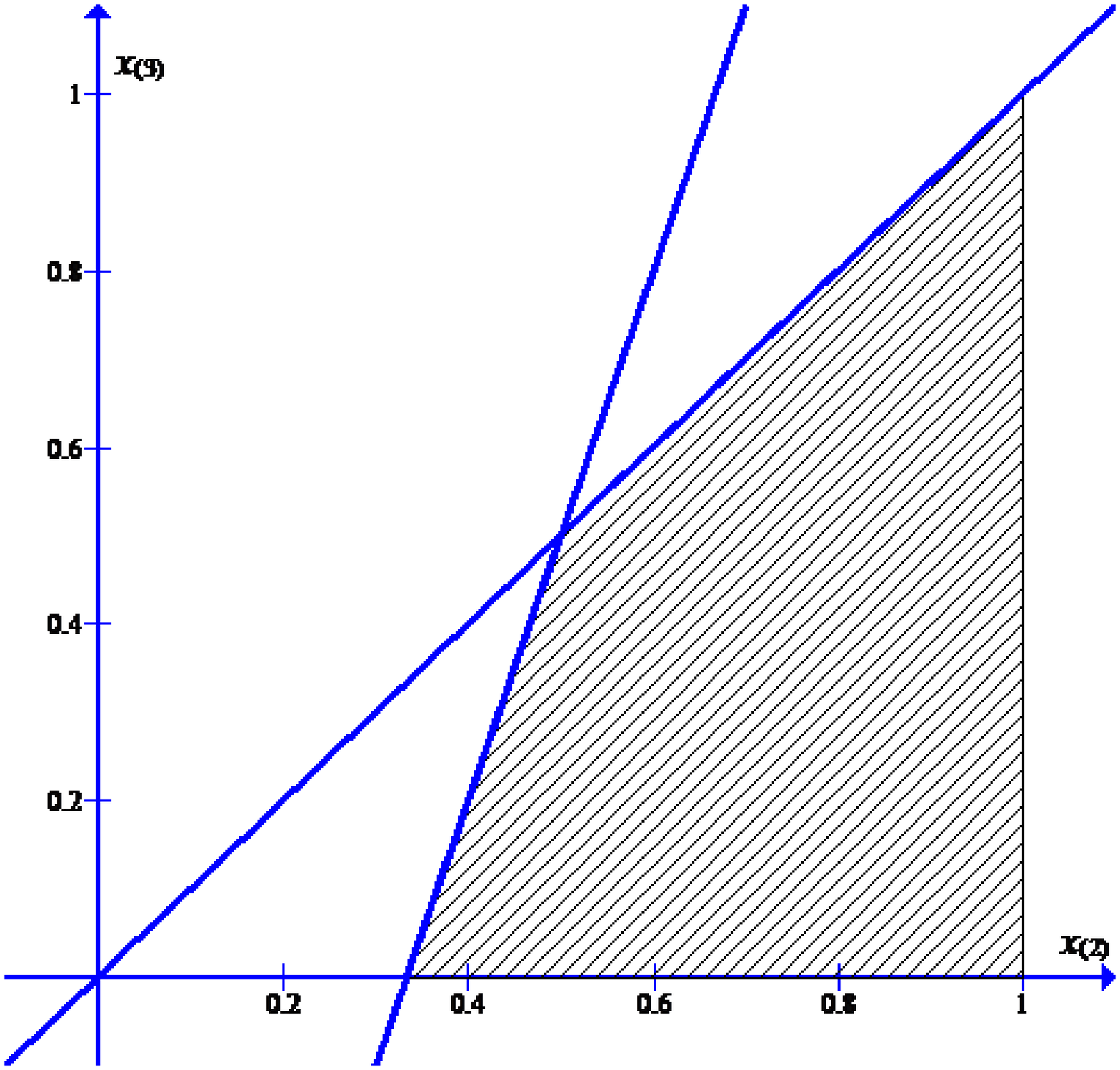}\caption{Allocation region.}
\label{fig:Allocation-region.}
\end{figure}

To illustrate the working of the optimal mechanism, suppose that there
are three buyers whose valuations are distributed uniformly between
zero and one. That is, $N=3$ and $F=U[0,1]$. In that case, virtual
valuations are given by $\psi(x)=2x-1$. The reserve rule is to allocate
when $3x_{(2)}-1>x_{(3)}.$ Then the good is always allocated when
$x_{(2)}\geq\psi^{-1}(0)=\frac{1}{2}$ and never allocated when $x_{(2)}<a(0)=\frac{1}{3}$.
Figure \ref{fig:Allocation-region.} illustrates the combinations
of values of $x_{(2)}$ and $x_{(3)}$ that lead to allocation.

What is the probability that the unit is allocated? To calculate that
probability, we integrate over the shaded area in Figure 1: 
\begin{equation}
1-F_{2}\left(\frac{1}{2}\right)+\int_{1/3}^{1/2}F_{3|x_{(2)}}(3x_{(2)}-1)f_{2}(x_{(2)}).\label{uniformprobability}
\end{equation}
Recall that $F_{2}$ is the distribution of the second order statistic
$X_{(2)}$ and $F_{3|x_{(2)}}$ is the distribution of $X_{(3)}$
conditional on $X_{(2)}=x_{(2)}.$ Here, $F_{2}(x)=3x^{2}-2x^{3},$
with density $f_{2}(x)=6x(1-x),$ and $F_{3|x_{(2)}}(x)=\frac{x}{x_{(2)}},$
with density $f_{3|x_{(2)}}(x)=\frac{1}{x_{(2)}}.$ Substituting these
definitions into Expression \ref{uniformprobability}, we find that
the probability of allocation is $\frac{23}{36}\approx0.64$.

Similarly, substituting $f_{2}$ and $f_{3|x_{(2)}}$ into Expression
\ref{expectedrevenue} gives the seller's expected revenue: 
\begin{align*}
 & {\int_{1/3}^{1/2}\int_{0}^{3x_{(2)}-1}\left[3x_{(2)}-1-x_{(3)}\right]}\frac{1}{x_{(2)}}{6x_{(2)}(1-x_{(2)})}\\
 & +{\int_{1/2}^{1}\int_{0}^{x_{(2)}}\left[3x_{(2)}-1-x_{(3)}\right]\frac{1}{x_{(2)}}6x_{(2)}(1-x_{(2)}).}
\end{align*}
Integrating these expressions yields expected revenue of $\frac{55}{144}\approx0.382$.
We can also compute the expected revenue to the second seller and
find that it is $\frac{125}{432}\approx0.289.$ By contrast, in the
absence of any reserve rule, both sellers earn $E[X_{(3)}]=0.25$.
Thus, the second seller also benefits when the first seller uses the
optimal reserve rule.

\section{The Optimal Mechanism when $r>\underline{x}$\label{sec:Positive reserve price copy(1)}}

We now consider the more general case where there may be a non-trivial
reserve price in the second auction ($r>\underline{x}$). We find
that whenever $r$ is below $\psi^{-1}(0)$, the optimal reserve price
in the standard setting, then the optimal mechanism for the first
seller is qualitatively very similar to what we found in the baseline
case of $r=0$. Differences arise only when exactly one or two bidders
have valuations above $r$, a scenario that is unlikely when $r$
is low or the number of bidders $N$ is large. When $r\geq\psi^{-1}(0)$,
then a mechanism closer to a standard auction is optimal for the first
seller.

Our analysis proceeds as follows. Now the payoff to a buyer $i$ with
valuation $X_{i}\geq r$ in the second period, provided that he did
not obtain the first object, depends not only on the seller's allocation
decision, but also on whether or not the highest and second highest
types among his rivals exceed $r$. As a result, we have more cases
to consider. The payoff to a buyer $i$ in the second period is
\[
\max\{X_{i}-\max\left\{ r,Y_{(2)}\right\} ,0\}
\]
if the first object is allocated to the competitor with the highest
type $Y_{(1)}$, and it is
\[
\max\left\{ X_{i}-\max\left\{ r,Y_{(1)}\right\} ,0\right\} 
\]
otherwise. Thus, the expected payoff to a bidder $i$ with type $x_{i}\geq r$,
given reports $\mathbf{x}$ and excluding any payment to the first
seller, is
\[
\begin{array}{c}
P_{i}(\mathbf{x)}\cdot x_{i}+P_{j(1)}(\mathbf{x)}\cdot\max\left\{ x_{i}-\max\left\{ r,y_{(2)}\right\} ,0\right\} \\
+\left(1-P_{i}(\mathbf{x)}-P_{j(1)}(\mathbf{x)}\right)\cdot\max\left\{ x_{i}-\max\left\{ r,y_{(1)}\right\} ,0\right\} ,
\end{array}
\]
where as before $y_{(1)}$ and $y_{(2)}$ denote the two highest valuations
among his competitors, $j(1)$ is the highest-type competitor (so
that $X_{j(1)}=y_{(1)}$), and $P_{i}(\mathbf{x)}$ denotes the probability
that the first object is assigned to buyer $i$ given $\mathbf{x.}$
The payoff to a buyer $i$ with type $x_{i}<r$ in the second period
is 0, so the expected payoff given $\mathbf{x}$ is just $P_{i}(\mathbf{x)}\cdot x_{i}.$

Based on these payoffs, the interim expected payoff to a buyer of
type $x\geq r$ when all buyers report truthfully, excluding any payment
to the first seller, can be expressed as
\[
\begin{array}{ccc}
\Pi(x|x) & = & {\int\limits _{\left[\underline{x},\bar{x}\right]^{N-1}:r>\widehat{y}_{1}}}\left(x-r+\hat{p}^{1}\left(\left(x;\widehat{\mathbf{y}}\right)\right)\cdot r\right)\widehat{\mathbf{g}}(\widehat{\mathbf{y}})\\
 & + & {\int\limits _{\left[\underline{x},\bar{x}\right]^{N-1}:x>\widehat{y}_{1}\geq r>\widehat{y}_{2}}}\left(x-\widehat{y}_{1}+\hat{p}^{1}\left(\left(x;\widehat{\mathbf{y}}\right)\right)\cdot\widehat{y}_{1}+\hat{p}^{2}\left(\left(x;\widehat{\mathbf{y}}\right)\right)\cdot\left[\widehat{y}_{1}-r\right]\right)\widehat{\mathbf{g}}(\widehat{\mathbf{y}})\\
 & + & {\int\limits _{\left[\underline{x},\bar{x}\right]^{N-1}:x>\widehat{y}_{1}\geq\widehat{y}_{2}\geq r}}\left(x-\widehat{y}_{1}+\hat{p}^{1}\left(\left(x;\widehat{\mathbf{y}}\right)\right)\cdot\widehat{y}_{1}+\hat{p}^{2}\left(\left(x;\widehat{\mathbf{y}}\right)\right)\cdot\left[\widehat{y}_{1}-\widehat{y}_{2}\right]\right)\widehat{\mathbf{g}}(\widehat{\mathbf{y}})\\
 & + & {\int\limits _{\left[\underline{x},\bar{x}\right]^{N-1}:\widehat{y}_{1}\geq x,r>\widehat{y}_{2}}}\left(\hat{p}^{1}\left(\left(x;\widehat{\mathbf{y}}\right)\right)\cdot\left[x-r\right]+\hat{p}^{2}\left(\left(x;\widehat{\mathbf{y}}\right)\right)\cdot x\right)\widehat{\mathbf{g}}(\widehat{\mathbf{y}})\\
 & + & {\int\limits _{\left[\underline{x},\bar{x}\right]^{N-1}:\widehat{y}_{1}\geq x>\widehat{y}_{2}\geq r}}\left(\hat{p}^{1}\left(\left(x;\widehat{\mathbf{y}}\right)\right)\cdot\left[x-\widehat{y}_{2}\right]+\hat{p}^{2}\left(\left(x;\widehat{\mathbf{y}}\right)\right)\cdot x\right)\widehat{\mathbf{g}}(\widehat{\mathbf{y}})\\
 & + & {\displaystyle \sum_{k=2}^{N-1}}\left[{\int\limits _{\left[\underline{x},\bar{x}\right]^{N-1}:\widehat{y}_{k}\geq x>\widehat{y}_{k+1}}}\left(\hat{p}^{k+1}\left(\left(x;\widehat{\mathbf{y}}\right)\right)\cdot x\right)\widehat{\mathbf{g}}(\widehat{\mathbf{y}})\right].
\end{array}
\]
The first three integral terms give the payoffs to the buyer when
his type $x$ is the highest, the fourth and fifth integral terms
are the payoffs when $x$ is the second highest type, and the last
term is the payoff when $x$ is the $k$-th highest type, $k\geq3$.
In each of these terms, if the first good is allocated to the buyer,
then his payoff is simply $x$. His payoff when he does not get the
first good depends on the values of the highest and second-highest
rival types and on whether or not the first good is allocated to one
of them. The six possible cases, beginning with the first term, are
as follows:\smallskip{}

\begin{itemize}
\item If all rivals have values below the reserve price (i.e., $r>\widehat{y}_{1}$),
then his continuation payoff is $x-r,$ regardless of whether or not
the first good is allocated to a rival.
\item If $x$ is the highest type, the highest rival type exceeds $r$,
and all other rival types are below $r$ (i.e., $x>\widehat{y}_{1}\geq r>\widehat{y}_{2})$,
then his continuation payoff is $x-r$ if the good is allocated to
the highest rival and $x-\widehat{y}_{1}$ if not.
\item If $x$ is the highest type and the two highest rival types both exceed
$r$ (i.e., $x>\widehat{y}_{1}$ and $\widehat{y}_{2}\geq r)$, then
his continuation payoff is $x-\widehat{y}_{2}$ if the good is allocated
to the highest rival and $x-\widehat{y}_{1}$ if not.
\item If the highest rival type exceeds $x$ and the second-highest rival
type is less than $r$ (i.e., $\widehat{y}_{1}\geq x$ and $\widehat{y}_{2}<r)$,
then his continuation payoff is $x-r$ if the good is allocated to
the highest rival and $0$ if not.
\item If $x$ lies between the highest and second highest rival types and
the second-highest rival type exceeds $r$ (i.e., $\widehat{y}_{1}\geq x>\widehat{y}_{2}\geq r)$,
then his continuation payoff is $x-\widehat{y}_{2}$ if the good is
allocated to the highest rival and $0$ if not.
\item If $x$ is less than the second-highest rival type (i.e., $\widehat{y}_{2}>x$),
then he can never win the second object, so his continuation payoff
is $0$.\smallskip{}
\end{itemize}
A buyer of type $x<r$ is also never going to win the second auction.
Therefore, his expected payoff from reporting truthfully is 
\begin{align*}
\Pi(x|x) & ={\int\limits _{\left[\underline{x},\bar{x}\right]^{N-1}:x>\widehat{y}_{1}}}\left(\hat{p}^{1}\left(\left(x;\widehat{\mathbf{y}}\right)\right)\cdot x\right)\widehat{\mathbf{g}}(\widehat{\mathbf{y}})\\
 & +{\displaystyle \sum_{k=1}^{N-1}}\left[{\int\limits _{\left[\underline{x},\bar{x}\right]^{N-1}:\widehat{y}_{k}\geq x>\widehat{y}_{k+1}}}\left(\hat{p}^{k+1}\left(\left(x;\widehat{\mathbf{y}}\right)\right)\cdot x\right)\widehat{\mathbf{g}}(\widehat{\mathbf{y}})\right].
\end{align*}

Proceeding as in Section \ref{sec:Optimal Mechanism}, we can show
that the expected transfer to the seller from each bidder is
\begin{align}
Et(\widehat{X}) & =t(\underline{x})-\Pi(\underline{x}|\underline{x})\label{eqrevenue-1}\\
 & +\frac{1}{N}\left\{ \begin{array}{cc}
 & {\int\limits _{\left[\underline{x},\bar{x}\right]^{N}:r>\widehat{x}_{1}}}\left(\hat{p}^{1}\left(\widehat{\mathbf{x}}\right)\cdot\psi(\widehat{x}_{1})\right)\mathbf{\hat{\mathbf{f}}}\left(\widehat{\mathbf{x}}\right)\\
+ & {\int\limits _{\left[\underline{x},\bar{x}\right]^{N}:\widehat{x}_{1}\geq r>\widehat{x}_{2}}}\left(\psi(\widehat{x}_{1})-r+\hat{p}^{1}(\widehat{\mathbf{x}})\cdot r\right)\mathbf{\hat{\mathbf{f}}}\left(\widehat{\mathbf{x}}\right)\\
+ & {\int\limits _{\left[\underline{x},\bar{x}\right]^{N}:\widehat{x}_{2}\geq r>\widehat{x}_{3}}}\left(\psi(\widehat{x}_{1})-\widehat{x}_{2}+\hat{p}^{1}(\widehat{\mathbf{x}})\cdot\widehat{x}_{2}+\hat{p}^{2}(\widehat{\mathbf{x}})\cdot\left[\widehat{x}_{2}-r\right]\right)\mathbf{\hat{\mathbf{f}}}\left(\widehat{\mathbf{x}}\right)\\
+ & {\int\limits _{\left[\underline{x},\bar{x}\right]^{N}:\widehat{x}_{3}\geq r}}\left(\psi(\widehat{x}_{1})-\widehat{x}_{2}+\hat{p}^{1}(\widehat{\mathbf{x}})\cdot\widehat{x}_{2}+\hat{p}^{2}(\widehat{\mathbf{x}})\cdot\left[\widehat{x}_{2}-\widehat{x}_{3}\right]\right)\mathbf{\hat{\mathbf{f}}}\left(\widehat{\mathbf{x}}\right)
\end{array}\right\} \nonumber \\
 & +\frac{1}{N}\left\{ \begin{array}{cc}
 & {\int\limits _{\left[\underline{x},\bar{x}\right]^{N}:r>\widehat{x}_{2}}}\left(\hat{p}^{2}\left(\widehat{\mathbf{x}}\right)\cdot\psi(\widehat{x}_{2})\right)\mathbf{\hat{\mathbf{f}}}\left(\widehat{\mathbf{x}}\right)\\
+ & {\int\limits _{\left[\underline{x},\bar{x}\right]^{N}:\widehat{x}_{2}\geq r>\widehat{x}_{3}}}\left(\hat{p}^{1}(\widehat{\mathbf{x}})\cdot\left[\psi\left(\widehat{x}_{2}\right)-r\right]+\hat{p}^{2}(\widehat{\mathbf{x}})\cdot\psi\left(\widehat{x}_{2}\right)\right)\mathbf{\hat{\mathbf{f}}}\left(\widehat{\mathbf{x}}\right)\\
+ & {\int\limits _{\left[\underline{x},\bar{x}\right]^{N}:\widehat{x}_{3}\geq r}}\left(\hat{p}^{1}(\widehat{\mathbf{x}})\cdot\left[\psi\left(\widehat{x}_{2}\right)-\widehat{x}_{3}\right]+\hat{p}^{2}(\widehat{\mathbf{x}})\cdot\psi\left(\widehat{x}_{2}\right)\right)\mathbf{\hat{\mathbf{f}}}\left(\widehat{\mathbf{x}}\right)
\end{array}\right\} \nonumber \\
 & +\frac{1}{N}\sum_{k=3}^{N}\int_{\left[\underline{x},\bar{x}\right]^{N}}\left(\hat{p}^{k}(\widehat{\mathbf{x}})\cdot\psi\left(\widehat{x}_{k}\right)\right)\mathbf{\hat{\mathbf{f}}}\left(\widehat{\mathbf{x}}\right).\nonumber 
\end{align}

The seller maximizes expected revenue $ER(\widehat{X})=N\cdot Et(\widehat{X})$
subject to incentive compatibility and individual rationality. Maximizing
that integral pointwise, as we did in Section \ref{sec:Optimal Mechanism},
yields a solution that may fail to be globally incentive compatible
(a buyer may prefer to report a type far from his own), as we will
see.

Given any vector of ordered types $\hat{\mathbf{x}}$, taking the
derivative of the seller's expected revenue with respect to $\hat{p}^{k}\left(\hat{\mathbf{x}}\right)$
yields the following:
\begin{enumerate}
\item If $\widehat{x}_{2}\geq r$, 
\[
\frac{\partial ER(\widehat{X})}{\delta\hat{p}^{1}\left(\hat{\mathbf{x}}\right)}=\frac{\partial ER(\widehat{X})}{\delta\hat{p}^{2}\left(\hat{\mathbf{x}}\right)}=\left[\psi\left(\widehat{x}_{2}\right)+\widehat{x}_{2}-\max\left\{ \widehat{x}_{3},r\right\} \right]\widehat{\mathbf{f}}(\widehat{\mathbf{x}}),
\]
and for all $k>2$, 
\[
\frac{\partial ER(\widehat{X})}{\delta\hat{p}^{k}\left(\hat{\mathbf{x}}\right)}=\psi\left(\widehat{x}_{k}\right)\widehat{\mathbf{f}}(\widehat{\mathbf{x}}).
\]
\item If $\widehat{x}_{1}\geq r>\widehat{x}_{2}$, 
\[
\frac{\partial ER(\widehat{X})}{\delta\hat{p}^{1}\left(\hat{\mathbf{x}}\right)}=r\widehat{\mathbf{f}}(\widehat{\mathbf{x}})
\]
and for all $k>1$, 
\[
\frac{\partial ER(\widehat{X})}{\delta\hat{p}^{k}\left(\hat{\mathbf{x}}\right)}=\psi\left(\widehat{x}_{k}\right)\widehat{\mathbf{f}}(\widehat{\mathbf{x}}).
\]
\item If $r>\widehat{x}_{1}$, for all $k$, 
\[
\frac{\partial ER(\widehat{X})}{\delta\hat{p}^{k}\left(\hat{\mathbf{x}}\right)}=\psi\left(\widehat{x}_{k}\right)\widehat{\mathbf{f}}(\widehat{\mathbf{x}}).
\]
\end{enumerate}
As before, global incentive compatibility is satisfied if a bidder
cannot increase his probability of getting an item (either the first
or the second) by underreporting his type, or decrease the probability
by overreporting his type. We find that the solution to this pointwise
maximization satisfies that condition when $r\in\left[\psi^{-1}(0),\bar{x}\right)$
but not when $r\in\left(\underline{x},\psi^{-1}(0)\right)$.

The source of the problem is a qualitative difference, relative to
the no-reserve case, in the marginal revenue expression when $\widehat{x}_{1}\geq r>\widehat{x}_{2}$.
If $\widehat{x}_{2}\geq r$, then the marginal revenue from allocating
to the highest or second-highest bidder, $\psi\left(\widehat{x}_{2}\right)+\widehat{x}_{2}-\max\left\{ \widehat{x}_{3},r\right\} $,
matches what it was in the $r=0$ case: the only change is that $\max\left\{ \widehat{x}_{3},r\right\} $
takes the place of $\widehat{x}_{3}$. If $\widehat{x}_{1}\geq r>\widehat{x}_{2}$,
however, then the marginal revenue from allocating to the highest
bidder is $r$, independent of the exact values of $\widehat{x}_{2}$
and $\widehat{x}_{3}$. As $\widehat{x}_{2}$ moves from just below
$r$ to just above $r$, marginal revenue jumps from strictly positive
to strictly negative. That downward switch drives the failure of global
incentive compatibility, which we explore below.

\subsection{If $r\in\left[\psi^{-1}(0),\bar{x}\right)$\label{subsec:If r>psi-1(0)}}

Recall that when the first seller does not face competition from a
second seller, then the optimal mechanism is to allocate the object
to the bidder with the highest valuation if and only if $\psi(\widehat{x}_{1})\geq0.$
Thus, if $r\in\left[\psi^{-1}(0),\bar{x}\right)$, then the second
seller is using a reserve price higher than the optimal reserve price
in the standard mechanism design setting. In this case, the solution
to the first seller's pointwise maximization problem is to
\begin{itemize}
\item allocate to the top bidder if $\widehat{x}_{1}\geq r>\widehat{x}_{2}$
or $r>\widehat{x}_{1}\geq\psi^{-1}(0)$, because the marginal revenue
for $\hat{p}^{1}$ is $r>0$ or $\psi\left(\widehat{x}_{1}\right)>0$,
respectively;
\item allocate to one of the top two bidders if $\widehat{x}_{2}\geq r$,
because the marginal revenue for both $\hat{p}^{1}$ and $\hat{p}^{2}$
is $\psi\left(\widehat{x}_{2}\right)+\widehat{x}_{2}-\max\left\{ \widehat{x}_{3},r\right\} \geq\psi\left(\widehat{x}_{2}\right)>0;$
\end{itemize}
and not to allocate otherwise. If the seller uses this rule (and any
method of breaking indifferences between allocating to the highest
and second-highest bidders), then it is straightforward to show that
the probability of getting an item (first or second) is increasing
in the report. Thus, global incentive compatibility is satisfied.

\begin{theorem} \label{r above 1/2}If the distribution of buyer
values $F$ has increasing virtual values and the reserve price in
the second auction is $r\in\left[\psi^{-1}(0),\bar{x}\right)$, then
the following is an optimal (direct) mechanism for the first seller.
(Ties are broken randomly.)
\begin{enumerate}
\item \textbf{\textit{Allocation rule:}}\textit{\ The seller allocates
the good if and only if} $\psi\left(x_{(1)}\right)\geq0;$\textit{
allocation in that case is to the bidder with the highest valuation
if $x_{(2)}<r$, and it is to either the bidder with the highest valuation
or the bidder with the second-highest valuation if} $x_{(2)}\geq r$\textit{.}
\item \textbf{\textit{Transfers:}}
\begin{enumerate}
\item \textit{If $\psi\left(x_{(1)}\right)\geq0\text{ and }x_{(2)}<r$,
then the bidder with the highest valuation pays $\max\left\{ \psi^{-1}(0),x_{(2)}\right\} $
and the other bidders pay nothing.}
\item \textit{If $x_{(2)}\geq r$, then the bidder who receives the object
pays $\max\left\{ r,x_{(3)}\right\} $ and the other bidders pay nothing.}
\item \textit{If the good is not allocated, then there are no payments.}\medskip{}
\end{enumerate}
\end{enumerate}
\end{theorem}

For brevity, we do not write out the expected revenues in Theorems
\ref{r above 1/2}, \ref{Theorem: optimal mechanism low r}, and \ref{Theorem: optimal mechanism high r}.
It is straightforward to calculate those expected revenues from the
specified transfer functions.

The mechanism in Theorem \ref{r above 1/2} can be implemented through
a hybrid second- and third-price auction with reserve prices: if the
highest bid is above $\psi^{-1}(0)$, then the item goes to the highest
bidder at a price equal to either 1) the maximum of $\psi^{-1}(0)$
and the second-highest bid if that second-highest bid is below $r$,
or 2) the maximum of $r$ and the third-highest bid if the second-highest
bid is above $r$.

\subsection{If $r\in\left(\underline{x},\psi^{-1}(0)\right)$}

In this case, the second seller uses a reserve price lower than the
optimal reserve price in the standard mechanism design setting, and
the solution to the pointwise maximization problem turns out to violate
incentive compatibility. Recall that for $r<\psi^{-1}(0)$, $a(r)\in\left[\underline{x},\bar{x}\right]$
is defined as the valuation that solves $a+\psi(a)=r.$ Note that
$a(r)>r$ and $a(r)<\psi^{-1}(0)$. Given this definition, the solution
to the pointwise maximization problem is to
\begin{itemize}
\item allocate to the top bidder if $\widehat{x}_{1}\geq r>\widehat{x}_{2}$;
\item allocate to one of the top two bidders if $\widehat{x}_{3}\geq r$
and $\widehat{x}_{2}+\psi\left(\widehat{x}_{2}\right)-\widehat{x}_{3}\geq0$;
\item allocate to one of the top two bidders if $\widehat{x}_{2}\geq a(r)$
and $r>\widehat{x}_{3}$, because the marginal revenue for both $\hat{p}^{1}$
and $\hat{p}^{2}$ is $\widehat{x}_{2}+\psi\left(\widehat{x}_{2}\right)-r\geq a(r)+\psi\left(a(r)\right)-r=0;$
\item not allocate if $a(r)>\widehat{x}_{2}$ and $r>\widehat{x}_{3}$,
because the marginal revenue for both $\hat{p}^{1}$ and $\hat{p}^{2}$
is $\widehat{x}_{2}+\psi\left(\widehat{x}_{2}\right)-r<a(r)+\psi\left(a(r)\right)-r=0;$
\end{itemize}
and not to allocate otherwise.

This rule is not incentive compatible. To see why, suppose that bidder
$i$ with type $x$ between $r$ and $a(r)$ considers deviating to
a report $x^{\prime}$ below $r$. If $x$ is the highest type, then
bidder $i$ is certain to get an item with either report: the first
item if he reports $x$, the second item if he reports $x^{\prime}$.
If $x$ is the third-highest or lower order type, then bidder $i$
will get nothing with either report. He will also get nothing if $x$
is the second-highest type and $\widehat{x}_{3}$ is greater than
$r$, because the first unit will not be allocated at either report
(the marginal revenue from doing so is negative), so bidder $i$ loses
the second auction to the bidder with the highest type. The non-monotonicity
arises when $x$ is the second-highest type and $\widehat{x}_{3}$
is less than $r$. If bidder $i$ reports truthfully, then the first
unit is not allocated and he loses the second auction to the bidder
with the highest type. But if he reports a type below $r,$ then the
rule above specifies that the highest bidder gets the first unit,
and then bidder $i$ will win the second. Thus, a bidder with a type
between $r$ and $a(r)$ is more likely to get an item by reporting
a type below $r$ than by reporting truthfully.

More formally, incentive compatibility requires the second-order condition
that
\begin{equation}
\int_{q}^{x}\Pi_{2}(x^{\prime}|x^{\prime})dx^{\prime}\geq\int_{q}^{x}\Pi_{2}(q|x^{\prime})dx^{\prime},\label{eq:global IC}
\end{equation}
for all $x,q\in\left[\underline{x},\bar{x}\right]$, where $\Pi_{2}(q|x)$,
the derivative of the gross payoff $\Pi(q|x)$ with respect to the
buyer's true type, corresponds to the probability that buyer of type
$x$ gets an item (either the first or the second) when reporting
type $q$. (See Section \ref{subsec:Incentive-compatibility}.) The
allocation rule derived above violates that condition at $x=r$ and
any $q<r$.

We proceed to find the optimal mechanism through a process of ``guess
and verify.\textquotedblright{} First, we guess that the constraints
in Expression \ref{eq:global IC} bind only for types $x$ below $a(r)$;
that for type $r$ the constraint binds only for underreports $q<r$;
and that for the rest of the types $x\leq a(r)$ the constraint binds
only for a marginal underreport, $q=x-\epsilon$ for vanishingly small
$\epsilon$.\footnote{Carroll and Segal \cite{CarrollSegal} encounter a similar failure
of the first-order approach to revenue maximization in an environment
with resale. Bergemann et al.\ \cite{BBM} also have to consider
non-local incentive compatibility. Briefly, the difference between
our approach to finding the optimal mechanism and the approaches in
those papers is that they start by guessing the mechanism and then
calculate the Lagrange multipliers on the incentive constraints that
turn out to bind, and we start by guessing which of the constraints
in Expression \ref{eq:global IC} bind and then calculate the corresponding
allocation rule.} Those guesses yield a continuum of constraints that take the form
\begin{equation}
\int_{q}^{r}\Pi_{2}(x^{\prime}|x^{\prime})dx^{\prime}\geq\int_{q}^{r}\Pi_{2}(q|x^{\prime})dx^{\prime}\label{eq:r downward constraint}
\end{equation}
for each $q\in\left[\underline{x},r\right)$ (let $\lambda_{r,q}$
denote the corresponding Lagrange multiplier); and
\begin{equation}
\Pi_{2}(x|x)\geq\limsup_{\epsilon\searrow0}\Pi_{2}(x-\epsilon|x)\label{eq:x immediate downward constraint}
\end{equation}
for each $x\in\left(\underline{x},a(r)\right]$ (multiplier $\mu_{x}$).
The constraints reflect the idea that a buyer must have a weakly higher
chance of getting an object if he reports truthfully than if he underreports.\footnote{Note that for type $r$, there is a redundancy: $\mu_{r}$ and $\lambda_{r,r-0}$
refer to the same constraint. In the proofs, it will be convenient
notationally to use one constraint in some cases and the other in
other cases.}

We derive first-order conditions by maximizing Expression \ref{eqrevenue-1}
subject to the constraints in Expressions \ref{eq:r downward constraint}
and \ref{eq:x immediate downward constraint}. We then guess the values
of $\lambda_{r,q}$ and $\mu_{x}$, derive the solution using those
guesses, and show that it satisfies incentive compatibility. (See
Appendix \ref{sec:Convexity-and-Incentive}.) The optimal mechanism
corresponds to pointwise maximization except when there are either
one or two bids above the reserve price $r$: the allocation rule
is the same as in the no-reserve case when there are at least three
bids above $r$, and it specifies no allocation when all bids are
below $r$. When $\widehat{x}_{1}>r>$ $\widehat{x}_{3}$ and $\widehat{x}_{2}$
is either below $r$ or just above it (between $r$ and $a(r)$),
then the constraints in Expressions \ref{eq:r downward constraint}
and \ref{eq:x immediate downward constraint} bind and the allocation
rule needs to be adjusted. Depending on the value of $r$, the solution
is to allocate either in all of these cases (regardless of the exact
valuations of the bidders) or in none of them.

The intuition for this solution is as follows. We would like to allocate
the good to the highest bidder when $\widehat{x}_{1}>r>\widehat{x}_{2}$,
because the marginal revenue $r$ from doing so is positive, but not
when $a(r)>\widehat{x}_{2}\geq r>\widehat{x}_{3}$, because in this
case the marginal revenue is negative (i.e., $\widehat{x}_{2}+\psi\left(\widehat{x}_{2}\right)-r<0).$
Roughly, the constraints in Expression \ref{eq:r downward constraint}
mean that if we allocate when $\widehat{x}_{1}=x^{*}$ and $\widehat{x}_{2}<r$,
then we also have to allocate when $\widehat{x}_{1}=x^{*}$ and $\widehat{x}_{2}=r$:
otherwise a bidder with type $r$ would be more likely to get an item
by underreporting. The constraints in Expression \ref{eq:x immediate downward constraint}
then imply that we must also allocate when $\widehat{x}_{2}$ is just
above $r$, and then when $\widehat{x}_{2}$ is just above that value,
and so on. Iterating those constraints, we conclude that if we allocate
when $\widehat{x}_{1}=x^{*}$ and $\widehat{x}_{2}<r$, then we must
also allocate when we replace $\widehat{x}_{2}$ with any higher value,
including values above $x^{*}$: that is, allocate whenever $\widehat{x}_{1}\geq x^{*}$.

The seller's maximization problem, then, consists of finding the optimal
cutoff $x^{*}$ such that when $\widehat{x}_{3}<r$ and $\widehat{x}_{2}<a(r)$,
the seller allocates if and only if $\widehat{x}_{1}\geq x^{*}$.
The resulting revenue is $N\cdot F\left(r\right)^{N-2}$ times $Z^{r}(x^{*})$,
where the function $Z^{r}(x^{*})$ is defined as follows:

\begin{definition} For $r\in\left[\underline{x},\bar{x}\right]$
and $x^{*}\in\left[r,\bar{x}\right]$, define
\[
Z^{r}(x^{*})\equiv rF\left(r\right)\left[1-F\left(x^{*}\right)\right]+(N-1)\int\limits _{x^{*}}^{\bar{x}}\left(\int\limits _{r}^{\min\left\{ x,a(r)\right\} }[\psi(x')+x'-r]f(x')dx'\right)f(x)dx.
\]
\end{definition}

To interpret $Z^{r}(x^{*})$, note that the revenue $N\cdot F\left(r\right)^{N-2}Z^{r}(x^{*})$
is the marginal revenue $r$ when $\widehat{x}_{1}\geq x^{*}$ and
$\widehat{x}_{2}<r$ times the probability of that event, plus the
(negative) expected marginal revenue when $\widehat{x}_{1}\geq x^{*}$
and $a(r)>\widehat{x}_{2}\geq r>\widehat{x}_{3}$ times the probability
of \emph{that} event. (Recall that when $\widehat{x}_{2}\geq a(r)$,
we want to allocate regardless, so we do not need to include that
case in the revenue maximization.) The function $Z^{r}(x^{*})$ is
quasiconvex, so the optimal $x^{*}$ is at a corner: either $x^{*}=r$
or $x^{*}=\bar{x}$. Observe that $Z^{r}(\bar{x})=0$ (because $\widehat{x}_{1}$
cannot exceed $\bar{x}$). Thus, $x^{*}=\bar{x}$ is optimal if and
only if $Z^{r}(r)\leq0$, because then revenue is higher at $x^{*}=\bar{x}$
than at $x^{*}=r$. If $Z(r)\geq0$, then $x^{*}=r$ is optimal. That
logic forms the basis for our guesses of the values of the Lagrange
multipliers.

The value of $Z^{r}(r)$ is decreasing in the number of bidders $N$.
When $N$ is large enough, all else equal, the optimal cutoff $x^{*}$
equals $\bar{x}$, and the item is not allocated unless the second-highest
bid is at least $a(r)$. We have not been able to establish whether
or not $Z^{r}(r)$ is monotonic in the reserve price $r$. We do know,
however, that for small enough values of $r$, again the optimal cutoff
$x^{*}$ equals $\bar{x}$. The reason is that $Z^{r}(r)$ is continuous
and strictly negative at $r=\underline{x}$. Thus, the optimal mechanism
changes continuously near the baseline case of no reserve price in
the second auction: for values of $r$ such that $Z^{r}(r)<0$, the
seller allocates if and only if\textit{
\begin{equation}
\psi\left(\widehat{x}_{2}\right)+\widehat{x}_{2}-\max\left\{ r,\widehat{x}_{3}\right\} \geq0.\label{eq:psi-x2 + x2- max r x3}
\end{equation}
}\textit{\emph{When $Z^{r}(r)>0$, then the seller allocates either
if Condition \ref{eq:psi-x2 + x2- max r x3} holds or if $\widehat{x}_{1}\geq r>\widehat{x}_{3}$.}}

We summarize the optimal mechanism in the following two theorems.

\begin{theorem} \label{Theorem: optimal mechanism low r}If the distribution
of buyer values $F$ has increasing virtual values, the reserve price
in the second auction is $r\in\left(\underline{x},\psi^{-1}(0)\right)$,
and $Z^{r}(r)\leq0$, then the following is an optimal (direct) mechanism
for the first seller. (Ties are broken randomly.)
\begin{enumerate}
\item \textbf{\textit{Allocation rule:}}\textit{\ The seller allocates
the good to the bidder with the second-highest valuation if} \textit{$\psi\left(\widehat{x}_{2}\right)+\widehat{x}_{2}-\max\left\{ r,\widehat{x}_{3}\right\} \geq0,$
and does not allocate otherwise.}
\item \textbf{\textit{Transfers:}}
\begin{enumerate}
\item \textit{If $x_{(3)}\leq r$ and the good is allocated ($x_{(2)}\geq a\left(r\right)$),
then the bidder with the highest valuation pays $a\left(r\right)-r>0$,
the bidder with the second-highest valuation pays $a\left(r\right)>0$,
and the other bidders pay nothing.}
\item \textit{If $x_{(3)}\in\left(r,\psi^{-1}(0)\right)$ and the good is
allocated ($x_{(2)}\geq a\left(x_{(3)}\right)$), then the bidder
with the highest valuation pays $a\left(x_{(3)}\right)-x_{(3)}>0$,
the bidder with the second-highest valuation pays $a\left(x_{(3)}\right)>0$,
and the other bidders pay nothing.}
\item \textit{If $\psi\left(x_{(3)}\right)\geq0$ (in which case the good
is allocated because $x_{(2)}\geq a\left(x_{(3)}\right)=x_{(3)}$),
then the bidder with the second-highest valuation pays $x_{(3)}>0$
and the other bidders pay nothing.}
\item \textit{If the good is not allocated, then there are no payments.}\medskip{}
\end{enumerate}
\end{enumerate}
\end{theorem}

\begin{theorem} \label{Theorem: optimal mechanism high r}If the
distribution of buyer values $F$ has increasing virtual values, the
reserve price in the second auction is $r\in\left(\underline{x},\psi^{-1}(0)\right)$,
and $Z^{r}(r)\geq0$, then the following is an optimal (direct) mechanism
for the first seller. (Ties are broken randomly.)
\begin{enumerate}
\item \textbf{\textit{Allocation rule:}}\textit{\ The seller allocates
to the bidder with the highest valuation if} $\widehat{x}_{1}\geq r>\widehat{x}_{2},$\textit{
allocates to either the bidder with the highest valuation or the bidder
with the second-highest valuation if} $\widehat{x}_{2}\geq r>\widehat{x}_{3},$
\textit{allocates to the bidder with the second-highest valuation
if} \textit{$\widehat{x}_{3}\geq r\text{ \emph{and }}\psi\left(\widehat{x}_{2}\right)+\widehat{x}_{2}-\widehat{x}_{3}\geq0,$
and does not allocate otherwise.}
\item \textbf{\textit{Transfers:}}
\begin{enumerate}
\item \textit{If $\widehat{x}_{1}\geq r>\widehat{x}_{2}$, then the bidder
with the highest valuation pays $r$ and the other bidders pay nothing.}
\item \textit{If $\widehat{x}_{2}\geq r>\widehat{x}_{3}$, then the bidder
who receives the object pays $r$ and the other bidders pay nothing.}
\item \textit{If $x_{(3)}\in\left(r,\psi^{-1}(0)\right)$ and the good is
allocated ($x_{(2)}\geq a\left(x_{(3)}\right)$), then the bidder
with the highest valuation pays $a\left(x_{(3)}\right)-x_{(3)}>0$,
the bidder with the second-highest valuation pays $a\left(x_{(3)}\right)>0$,
and the other bidders pay nothing.}
\item \textit{If $\psi\left(x_{(3)}\right)\geq0$ (in which case the good
is allocated because $x_{(2)}\geq a\left(x_{(3)}\right)=x_{(3)}$),
then the bidder with the second-highest valuation pays $x_{(3)}>0$
and the other bidders pay nothing.}
\item \textit{If the good is not allocated, then there are no payments.}\medskip{}
\end{enumerate}
\end{enumerate}
\end{theorem}

This mechanism is qualitatively similar to the optimal mechanism for
the baseline case of no reserve price in the second auction, but it
has some interesting new features. The optimal withholding rule now
is a function of the first-, second-, and third-highest values, rather
than just the second- and third-highest. Further, the withholding
rule now varies with the number of bidders, unlike both our baseline
case and the standard auction environment: as mentioned above, $Z^{r}(r)$
is decreasing in $N$.

\section{Implementing the Optimal Mechanism\label{sec:Implementation}}

In this section, we show that the optimal mechanism can be implemented
either with a modified third-price auction or with a modified pay-your-bid
auction featuring a rebate. For simplicity, we focus on the case where
there is no reserve price in the second auction ($r=0$), but the
arguments extend to the case of a non-trivial $r$.

It is straightforward to implement the payments and allocation rule
from Theorem \ref{Theorem: optimal mechanism} in a version of a third-price
auction. Define the \emph{modified third-price auction} as follows:
each buyer submits a bid in $\left[\underline{x},\bar{x}\right]$.
As a function of the vector of bids $\mathbf{b}$, the good is allocated
to the second-highest bidder if and only if $b_{(2)}\geq a(b_{(3)}).$
If the unit is not allocated, then no one makes any payments. If the
unit is allocated, then the payments are based on the third-highest
bid, $b_{(3)}.$ When $\psi(b_{(3)})>0$, the highest bidder pays
nothing and the second-highest bidder pays $b_{(3)}$; when $\psi(b_{(3)})<0$,
then the highest bidder pays $a(b_{(3)})-b_{(3)}>0$ and the second-highest
bidder pays $a(b_{(3)})$.

\begin{theorem} If the distribution of buyer values $F$ has increasing
virtual valuations, then truthful bidding is an equilibrium of the
modified third-price auction, and that equilibrium yields the optimal
expected revenue for the first seller. \end{theorem}

In fact, truthful reporting is an ex post equilibrium. Consider, for
example, the highest-valuation buyer in the case where $\psi(x_{(3)})<0$
and the item is allocated ($x_{(2)}\geq a(x_{(3)})$). Truthfully
bidding $b=x_{(1)}$ yields a payoff of
\[
x_{(1)}-x_{(3)}-\left[a(x_{(3)})-x_{(3)}\right]=x_{(1)}-a(x_{(3)});
\]
the bidder transfers $a(x_{(3)})-x_{(3)})$ to the first seller and
then wins the second auction at price $x_{(3)}$. Any bid above $x_{(2)}$
yields that same payoff. A bid between $a(x_{(3)})$ and $x_{(2)}$
also results in payoff $x_{(1)}-a(x_{(3)})$: the bidder gets the
first item and transfers $a(x_{(3)})$ to the first seller. Any bid
below $a(x_{(3)})$ gives a lower payoff, $x_{(1)}-x_{(2)}$, because
the first item will not be allocated, no transfers will be made to
the first seller, and the bidder will win the second item at price
$x_{(2)}$. The other cases are similar.

Another way to implement the optimal mechanism is with a modified
first-price or ``pay your bid\textquotedblright{} auction, although
the construction is more complicated. In Theorem \ref{Theorem: optimal mechanism},
a bidder's payment depends on whether his is the highest or second-highest
bid, but he submits only a single bid. One solution is to implement
the highest bidder's transfer as an \emph{unconditional} (i.e., regardless
of whether or not the good is allocated) payment together with a \emph{rebate}
equal to the winning price in the second auction.

More formally, we show in Online Appendix \ref{sec:Modified-Pay-Your-Bid-Auction}
that a \emph{pay-your-bid auction} with the following rules implements
the optimal mechanism. A buyer with valuation $x$ submits a bid of
$\beta(x)$. The seller allocates the object according to Expression
\ref{eq:Reserve Rule}, the rule from the optimal mechanism. It will
turn out that $\beta(\cdot)$ is strictly increasing, so the seller
can implement that rule. If the item is allocated, then both the highest
and second-highest bidders pay their bids. If the item is not allocated,
then only the highest bidder pays his bid. In either case, the highest
bidder then gets a rebate equal to the sale price of the second item
($x_{(3)}$ if the first item is allocated, $x_{(2)}$ if it is not)
assuming that he wins the second auction. The highest bidder does
not get a rebate if he does not win the second auction.

\section{Revenue Comparisons\label{sec:Revenues}}

In this section, we compare the expected revenue of the optimal mechanism
to the expected revenue of a standard auction with an optimal reserve
price. Relative to an optimal standard auction, how much better does
the optimal mechanism with its more complicated reserve rule do? We
show that the gains can be substantial. As in Section \ref{sec:Implementation},
for simplicity we study the case with no reserve price in the second
auction ($r=0$). The first step is derive the symmetric equilibrium
in the first auction with a positive reserve price, $r_{1}$. This
derivation turns out to be a significant challenge, because the equilibrium
involves pooling.

\begin{proposition} When $r=0$, then for any $r_{1}\in(0,E[Y_{(1)}])$,
there is no strictly increasing, symmetric pure-strategy equilibrium
of either a first-price auction or a second-price auction with reserve
price $r_{1}$ for the first good. \end{proposition}

(If the reserve price $r_{1}$ exceeds $E[Y_{(1)}]$, the expectation
of the highest rival value, then no one will submit a bid above the
reserve price in the first auction.) To see the reasoning behind the
non-existence result, consider a second-price auction with reserve
price $r_{1}$ and suppose that there is a symmetric equilibrium with
a strictly increasing bidding function $\beta$. The first-order condition
for an optimal bid above $r_{1}$ gives
\begin{equation}
\beta(x)=E[Y_{(2)}|Y_{(1)}=x],\label{eq:eqm beta}
\end{equation}
the expected price in the second auction conditional on losing the
first auction to another bidder of type $x$. Let $\hat{x}\geq r_{1}$
denote the lowest valuation such that a buyer submits a bid. A buyer
with valuation $\hat{x}$ must be indifferent between submitting a
bid of $\beta(\hat{x})$ in the first auction and not submitting a
bid. (If he strictly preferred to bid, then so would nearby types,
and $\hat{x}$ would not be the lowest type to submit a bid.) The
expected total payoff from submitting $\beta(\hat{x})$ is 
\[
(\hat{x}-r)G_{1}(\hat{x})+\int_{\hat{x}}^{\bar{x}}\left[\int_{\underline{x}}^{\hat{x}}[\hat{x}-y_{(2)}]g_{2|y_{(1)}}(y_{(2)})\right]g_{1}(y_{(1)});
\]
the expected payoff from not submitting a bid is 
\[
\int_{\underline{x}}^{\hat{x}}[x-y_{(1)}]g_{1}(y_{(1)})+\int_{\hat{x}}^{\bar{x}}\left[\int_{\underline{x}}^{\hat{x}}[\hat{x}-y_{(2)}]g_{2|y_{(1)}}(y_{(2)})\right]g_{1}(y_{(1)}).
\]
The difference is zero when 
\[
(\hat{x}-r_{1})G_{1}(\hat{x})=\int_{\underline{x}}^{\hat{x}}[x-y_{(1)}]g_{1}(y_{(1)})\Leftrightarrow r_{1}=\frac{1}{G_{1}(\hat{x})}\int_{\underline{x}}^{\hat{x}}y_{(1)}g_{1}(y_{(1)})=E[Y_{(1)}|Y_{(1)}\leq\hat{x}].
\]
But that value of $r_{1}$ is strictly greater than the value of $\beta(\hat{x})$
from Expression \ref{eq:eqm beta}. (The former is the expectation
of the highest of $N-1$ valuations, conditional on all being below
$\hat{x}$, while the latter is the expectation of the highest of
$N-2$, again conditional on all being below $\hat{x}$.) Thus, these
two necessary conditions for equilibrium are incompatible, and we
conclude that no strictly increasing, symmetric equilibrium of the
second-price auction with reserve price $r_{1}$ exists.

The non-existence result is not surprising. In our model, allocating
the good in the first auction generates a positive externality for
the losing buyers. Jehiel and Moldovanu \cite{JehielMoldovanu} were
the first to observe that a pure-strategy symmetric separating equilibrium
does not exist in a second-price auction with positive externalities.
However, they show that a symmetric equilibrium with partial pooling
at the reserve price can exist. In that equilibrium, an interval of
types $[\hat{x},\hat{\hat{x}}]$ all bid $r_{1}$, types above $\hat{\hat{x}}$
bid according to a strictly increasing $\beta(x)$, and types below
$\hat{x}$ do not bid. We construct such an equilibrium for our example
and then calculate the optimal reserve price and revenues for the
first seller.

\subsection{Three bidders, uniform valuations}

We will derive the partial-pooling equilibrium of the second price
auction with reserve $r_{1}$ for the $N=3$, $F=U[0,1]$, $r=0$
case and compute the optimal reserve price of the first seller and
associated revenue. Details are in Online Appendix \ref{sec:Thresholds-in-Second-Price}.
In the example, Expression \ref{eq:eqm beta} becomes $\beta(x)=x/2$.
The cutoff values $\hat{x}$ and $\hat{\hat{x}}$ are characterized
by two indifference conditions. A buyer of type $\hat{\hat{x}}$ is
indifferent between bidding $r_{1}$ (and tying with other types in
$[\hat{x},\hat{\hat{x}}]$) and bidding just above $r_{1}$; a buyer
of type $\hat{x}$ is indifferent between bidding $r$ and not bidding.
The type-$\hat{\hat{x}}$ buyer trades off overpaying for the first
item relative to the expected price in the second auction when there
is only one rival with a type in $[\hat{x},\hat{\hat{x}}]$ against
underpaying when there are two such rivals. The type-$\hat{x}$ buyer
overpays when there are 0 or 1 rival with type in $[\hat{x},\hat{\hat{x}}]$,
but may get an item even when both rivals have higher types.

Solving the two indifference conditions gives $\hat{x}=(1+1/\sqrt{3})r_{1}$
and $\hat{\hat{x}}=(1+2/\sqrt{3})r_{1}$. We next find the optimal
reserve price $r_{1}^{\ast}$ by maximizing the seller's expected
revenue $R_{1}(r_{1})$:
\[
R_{1}(r_{1})=\left[F_{1}(\hat{\hat{x}})-F_{1}(\hat{x})\right]r_{1}+\int\limits _{\hat{\hat{x}}}^{1}\left[F_{2|x_{(1)}}(\hat{\hat{x}})r_{1}+\int\limits _{\hat{\hat{x}}}^{x_{(1)}}\beta(x_{(2)})f_{2|x_{(1)}}(x_{(2)})\right]f_{1}(x_{(1)}).
\]
The solution is $r_{1}^{\ast}\approx0.379$. The corresponding values
of $\hat{x}$ and $\hat{\hat{x}}$ are $\hat{x}\approx0.60$ and $\hat{\hat{x}}\approx0.82$,
and the resulting maximal revenue is $R_{1}(r_{1}^{\ast})\approx0.303.$

The revenue of the second seller is the second-highest valuation $x_{(2)}$
if the first seller does not allocate and the third-highest valuation
$x_{(3)}$ otherwise -- except if all three valuations are between
$\hat{x}$ and $\hat{\hat{x}}$ and the first seller randomly allocates
to the buyer with valuation $x_{(3)}$, in which case the second seller
gets $x_{(2)}$ instead of $x_{(3)}$. Overall, the expected revenue
for the second seller when the first seller sets the optimal reserve
price is $R_{2}(r_{1}^{\ast})\approx0.282.$

A striking feature of this equilibrium is that the threshold for bidding,
$\hat{x},$ is significantly higher than the optimal reserve price,
$r_{1}^{\ast}$. The outside option of winning the second auction
at a price below $r_{1}^{\ast}$ causes types between $r_{1}^{\ast}$
and $\hat{x}$ not to bid in the first auction. Their lack of participation
gives the high types an incentive to participate because they are
more likely to win the first auction at price equal to $r_{1}^{\ast}$.
As a result, only 40\% of the buyers bid in the first auction and
roughly half of them bid the reserve price.

Table 1 summarizes the revenue results for our uniform example. 
\[
\begin{tabular}{|l|l|l|}
\hline \multicolumn{3}{|c|}{\textbf{Table 1: Revenue Comparisons when }\ensuremath{r=0}}\\
\hline   &  First Seller Revenues  &  Second Seller Revenues\\
\hline  Optimal Mechanism  & \multicolumn{1}{|c|}{0.382 } & \multicolumn{1}{|c|}{0.289}\\
\hline  Must-Sell Mechanism  & \multicolumn{1}{|c|}{0.250 } & \multicolumn{1}{|c|}{0.250}\\
\hline  Optimal Second Price Auction  & \multicolumn{1}{|c|}{0.303 } & \multicolumn{1}{|c|}{0.282 }
\\\hline \end{tabular}\ \ 
\]

\noindent In comparison to the must-sell mechanism, the optimal mechanism
increases the expected revenues of both sellers, by 54\% for the first
seller and by 16\% for the second seller. A standard auction with
an optimal reserve price gives the second seller essentially the same
increase but gives the first seller only a 20\% increase in revenues.

Table 1 compares revenues only for the case where the second seller
does not use a reserve price. However, the above analysis strongly
suggests that reserve prices in standard auctions are not a very effective
way for the first seller to increase revenues in a sequential auction
setting.

\section{Extensions\label{sec:Extensions}}

The design of the optimal mechanism can be straightforwardly extended
to environments in which either or both sellers have multiple units.
For example, suppose that the first seller has one unit to sell and
the second seller has $M$ units that she sells simultaneously in
a uniform-price auction with no reserve. Buyers have a weakly dominant
strategy to bid their value in second period so there is no leakage
problem. Then the optimal allocation rule is to allocate the unit
to the $\left(M+1\right)$-th highest type if and only if $\psi(x_{(M+1)})+M(x_{(M+1)}-x_{(M+2)})>0$.
The first term is the virtual valuation of the marginal buyer (the
one who gets an object if the first seller allocates and not otherwise),
and the second term is the total savings to the $M$ buyers from reducing
the price in the second auction from $x_{\left(M+1\right)}$ to $x_{\left(M+2\right)}$.

More broadly, we hope that our analysis can be used as a framework
for studying sequential competition in mechanisms between sellers,
in a setting where the second seller also acts strategically in choosing
her mechanism.\footnote{As a ``proof of concept,'' we have taken a first step in that direction
by analyzing a simplified form of competition, where the second seller
chooses her reserve price knowing that the first seller will respond
with an optimal mechanism. We find that in equilibrium, the second
seller's reserve price is below $\psi^{-1}(0)$, and consequently
the first seller uses a withholding rule that cannot be implemented
with a simple reserve price. In the $N=3$, $F=U[0,1]$ example, we
calculate that in equilibrium the second seller uses a reserve price
to increase her revenue at the expense of the first seller. Details
are available upon request.} Although in many real-world settings the relevant competition facing
a seller comes from future auctions (or, equivalently, from a Bertrand
market), we believe that such an extension would be valuable.For instance,
the optimal response of the second seller may not be a second-price
auction with a reserve price. When the first unit is allocated to
the second-highest type, then the second seller faces a distribution
of types that may not have increasing virtual valuations, even if
the original distribution does. The inherited distribution is ``hollowed
out,\textquotedblright{} in the sense that a middle value is the one
that gets removed. Further, the first seller's optimal allocation
rule implies that the types of the bidders remaining to face the second
seller are correlated.

The spillover effect between sequential sellers that we have identified
is conceptually distinct from the problem of information leakage,
but the two issues may interact. In general, the best response for
the second seller depends on what information about the bidders' types
is disclosed after the first period. Consider the extreme case (as
in Carroll and Segal \cite{CarrollSegal}) where all private information
is exogenously revealed after the first mechanism is run. If the second
seller has all the bargaining power, then she will make a take-it-or-leave-it
offer to the highest-type remaining buyer at exactly his value. Since
the buyers anticipate that they will get no surplus in the second
period, the problem facing the first seller is equivalent to the standard
mechanism design environment. On the other hand, if the buyers have
all the bargaining power in the second period, then the remaining
buyer with the highest value will get the second item at a price equal
to the second-highest remaining value, and so our optimal mechanism
emerges as the equilibrium choice for the first seller.

Information leakage may similarly complicate outcomes in the case
of more than two sequential sellers, if buyers worry that information
about their types revealed in their bids will influence the future
bids of their competitors. One potential solution would be to restrict
sellers to ex post incentive compatible mechanisms, in order to isolate
the effects of sequential competition in mechanisms from those of
information leakage.

\section{Concluding Remarks\label{sec:Conclusions}}

In sequential auction environments, losers of one auction can try
to buy again, typically from a different seller. In this paper we
show that for a seller who faces such competition from a subsequent
auction, using a standard first- or second-price auction with a reserve
price does not maximize revenue. Instead, we characterize the optimal
mechanism for any given reserve price by the second seller, using
techniques that may be useful in other settings where the first-order
approach does not yield an incentive compatible solution. The optimal
mechanism features payments from the top two bidders and a reserve
rule that depends on the two or three highest valuations. We also
present a third-price auction and a pay-your-bid auction with a rebate
that can be used in practice to implement the optimal mechanism.

We formulate our model in terms of sale auctions, but we can equally
interpret it as a model of procurement auctions, where the bidders
are potential sellers and their types represent their production costs.
An important motivating example is the market for pharmaceuticals
in Ecuador and other middle-income countries, studied by Brugués \cite{Brugues}.
There, the first buyer is the government, who procures a supply for
the public market. Losing bidders compete to serve the private market,
where Bertrand competition yields the same outcome as a second-price
auction with no reserve. Our analysis may be especially relevant in
this environment, because a government agency may have greater ability
to implement non-standard auction rules than would a private seller.

Our setting is a special case of a mechanism design environment with
externalities: when a bidder wins the first auction, then he will
not compete in the second auction. His absence, if he has the highest
or second-highest type, increases the continuation payoff (that is,
the payoff from the second auction) for the bidder with the highest
remaining type. We analyze how a seller can increase revenues or a
buyer can reduce her procurement payments by accounting for this externality
in the design of her auction.

We view our analysis as a first step in studying competition in mechanisms
between sellers in a sequential setting. There are many ways of modeling
such competition with regard to the timing of moves and information
revelation. One interesting possibility is to assume that types are
revealed after the first seller runs her auction (as is done in Bergemann
et al.\ \cite{BBM} and in Carroll and Segal \cite{CarrollSegal})
and to model the second stage as a Nash bargaining game between the
second seller and the remaining buyer with the highest value. The
Nash bargaining solution breaks the tie between allocating to the
highest or second-highest bidder in the optimal mechanism for the
first seller -- the surplus from allocating to the highest bidder
strictly strictly exceeds the surplus from allocating to the second-highest
bidder. As a result, the pointwise revenue maximizing solution may
fail to be incentive compatible.

\section{Appendix: Proving Theorems \ref{r above 1/2}, \ref{Theorem: optimal mechanism low r},
and \ref{Theorem: optimal mechanism high r}\label{sec:Convexity-and-Incentive}}

\subsection{Payoff from false report}

We derive the payoff to a type-$x$ buyer who reports his type as
$q$. If $x\geq r$ and $q\geq x$, then
\[
\begin{array}{ccc}
\Pi(q|x) & = & {\int\limits _{\left[\underline{x},\bar{x}\right]^{N-1}:x>\widehat{y}_{1}}}\left(\begin{array}{c}
x-\max\left\{ \widehat{y}_{1},r\right\} +\hat{p}^{1}\left(\left(q;\widehat{\mathbf{y}}\right)\right)\cdot\max\left\{ \widehat{y}_{1},r\right\} \\
+\hat{p}^{2}\left(\left(q;\widehat{\mathbf{y}}\right)\right)\cdot\left[\max\left\{ \widehat{y}_{1},r\right\} -\max\left\{ \widehat{y}_{2},r\right\} \right]
\end{array}\right)\widehat{\mathbf{g}}(\widehat{\mathbf{y}})\\
 & + & {\int\limits _{\left[\underline{x},\bar{x}\right]^{N-1}:q>\widehat{y}_{1}\geq x>\widehat{y}_{2}}}\left(\hat{p}^{1}\left(\left(q;\widehat{\mathbf{y}}\right)\right)\cdot x+\hat{p}^{2}\left(\left(q;\widehat{\mathbf{y}}\right)\right)\cdot\left[x-\max\left\{ \widehat{y}_{2},r\right\} \right]\right)\widehat{\mathbf{g}}(\widehat{\mathbf{y}})\\
 & + & {\int\limits _{\left[\underline{x},\bar{x}\right]^{N-1}:q>\widehat{y}_{1}>\widehat{y}_{2}\geq x}}\left(\hat{p}^{1}\left(\left(q;\widehat{\mathbf{y}}\right)\right)\cdot x\right)\widehat{\mathbf{g}}(\widehat{\mathbf{y}})\\
 & + & {\int\limits _{\left[\underline{x},\bar{x}\right]^{N-1}:\widehat{y}_{1}\geq q\geq x>\widehat{y}_{2}}}\left(\hat{p}^{1}\left(\left(q;\widehat{\mathbf{y}}\right)\right)\cdot\left[x-\max\left\{ \widehat{y}_{2},r\right\} \right]+\hat{p}^{2}\left(\left(q;\widehat{\mathbf{y}}\right)\right)\cdot x\right)\widehat{\mathbf{g}}(\widehat{\mathbf{y}})\\
 & + & {\displaystyle \sum_{k=1}^{N-1}}\left[{\int\limits _{\left[\underline{x},\bar{x}\right]^{N-1}:\widehat{y}_{k}\geq q>\widehat{y}_{k+1},\widehat{y}_{2}\geq x}}\left(\hat{p}^{k+1}\left(\left(q;\widehat{\mathbf{y}}\right)\right)\cdot x\right)\widehat{\mathbf{g}}(\widehat{\mathbf{y}})\right].
\end{array}
\]
If $x\geq r$ and $q<x$, then
\[
\begin{array}{ccc}
\Pi(q|x) & = & {\int\limits _{\left[\underline{x},\bar{x}\right]^{N-1}:q>\widehat{y}_{1}}}\left(\begin{array}{c}
x-\max\left\{ \widehat{y}_{1},r\right\} +\hat{p}^{1}\left(\left(q;\widehat{\mathbf{y}}\right)\right)\cdot\max\left\{ \widehat{y}_{1},r\right\} \\
+\hat{p}^{2}\left(\left(q;\widehat{\mathbf{y}}\right)\right)\cdot\left[\max\left\{ \widehat{y}_{1},r\right\} -\max\left\{ \widehat{y}_{2},r\right\} \right]
\end{array}\right)\widehat{\mathbf{g}}(\widehat{\mathbf{y}})\\
 & + & {\displaystyle \sum_{k=1}^{N-1}}\left[{\int\limits _{\left[\underline{x},\bar{x}\right]^{N-1}:\widehat{y}_{k}\geq q>\widehat{y}_{k+1},x>\widehat{y}_{1}}}\left(\begin{array}{c}
x-\max\left\{ \widehat{y}_{1},r\right\} \\
+\hat{p}^{1}\left(\left(q;\widehat{\mathbf{y}}\right)\right)\cdot\left[\max\left\{ \widehat{y}_{1},r\right\} -\max\left\{ \widehat{y}_{2},r\right\} \right]\\
+\hat{p}^{k+1}\left(\left(q;\widehat{\mathbf{y}}\right)\right)\cdot\max\left\{ \widehat{y}_{1},r\right\} 
\end{array}\right)\widehat{\mathbf{g}}(\widehat{\mathbf{y}})\right]\\
 & + & {\displaystyle \sum_{k=1}^{N-1}}\left[{\int\limits _{\left[\underline{x},\bar{x}\right]^{N-1}:\widehat{y}_{k}\geq q>\widehat{y}_{k+1},\widehat{y}_{1}\geq x>\widehat{y}_{2}}}\left(\begin{array}{c}
\hat{p}^{1}\left(\left(q;\widehat{\mathbf{y}}\right)\right)\cdot\left[x-\max\left\{ \widehat{y}_{2},r\right\} \right]\\
+\hat{p}^{k+1}\left(\left(q;\widehat{\mathbf{y}}\right)\right)\cdot x
\end{array}\right)\widehat{\mathbf{g}}(\widehat{\mathbf{y}})\right]\\
 & + & {\displaystyle \sum_{k=2}^{N-1}}\left[{\int\limits _{\left[\underline{x},\bar{x}\right]^{N-1}:\widehat{y}_{k}\geq q>\widehat{y}_{k+1},\widehat{y}_{2}\geq x}}\left(\hat{p}^{k+1}\left(\left(q;\widehat{\mathbf{y}}\right)\right)\cdot x\right)\widehat{\mathbf{g}}(\widehat{\mathbf{y}})\right].
\end{array}
\]
If $x<r$, then
\[
\Pi(q|x)={\int\limits _{\left[\underline{x},\bar{x}\right]^{N-1}:q>\widehat{y}_{1}}}\left(\hat{p}^{1}\left(\left(q;\widehat{\mathbf{y}}\right)\right)\cdot x\right)\widehat{\mathbf{g}}(\widehat{\mathbf{y}})+{\displaystyle \sum_{k=1}^{N-1}}\left[{\int\limits _{\left[\underline{x},\bar{x}\right]^{N-1}:\widehat{y}_{k}\geq q>\widehat{y}_{k+1}}}\left(\hat{p}^{k+1}\left(\left(q;\widehat{\mathbf{y}}\right)\right)\cdot x\right)\widehat{\mathbf{g}}(\widehat{\mathbf{y}})\right].
\]

\subsubsection{$\Pi_{2}(q|x)$\label{subsec:Pi-2}}

The derivative of the payoff with respect to its second argument (the
buyer's true type), $\Pi_{2}(q|x)$, will be used below. If $x\geq r$
and $q\geq x$, then we calculate that derivative as
\[
\begin{array}{ccc}
\Pi_{2}(q|x) & = & {\int\limits _{\left[\underline{x},\bar{x}\right]^{N-1}:x>\widehat{y}_{1}}}\widehat{\mathbf{g}}(\widehat{\mathbf{y}})+{\int\limits _{\left[\underline{x},\bar{x}\right]^{N-1}:\widehat{y}_{1}\geq x>\widehat{y}_{2}}}\left(\hat{p}^{1}\left(\left(q;\widehat{\mathbf{y}}\right)\right)+\hat{p}^{2}\left(\left(q;\widehat{\mathbf{y}}\right)\right)\right)\widehat{\mathbf{g}}(\widehat{\mathbf{y}})\\
 & + & {\int\limits _{\left[\underline{x},\bar{x}\right]^{N-1}:q>\widehat{y}_{1}>\widehat{y}_{2}\geq x}}\left(\hat{p}^{1}\left(\left(q;\widehat{\mathbf{y}}\right)\right)\right)\widehat{\mathbf{g}}(\widehat{\mathbf{y}})\\
 & + & {\displaystyle \sum_{k=1}^{N-1}}\left[{\int\limits _{\left[\underline{x},\bar{x}\right]^{N-1}:\widehat{y}_{k}\geq q>\widehat{y}_{k+1},\widehat{y}_{2}\geq x}}\left(\hat{p}^{k+1}\left(\left(q;\widehat{\mathbf{y}}\right)\right)\right)\widehat{\mathbf{g}}(\widehat{\mathbf{y}})\right].
\end{array}
\]
If $x\geq r$ and $q<x$, then
\[
\begin{array}{ccc}
\Pi_{2}(q|x) & = & {\int\limits _{\left[\underline{x},\bar{x}\right]^{N-1}:x>\widehat{y}_{1}}}\widehat{\mathbf{g}}(\widehat{\mathbf{y}})+{\displaystyle \sum_{k=1}^{N-1}}\left[{\int\limits _{\left[\underline{x},\bar{x}\right]^{N-1}:\widehat{y}_{k}\geq q>\widehat{y}_{k+1},\widehat{y}_{1}\geq x>\widehat{y}_{2}}}\left(\begin{array}{c}
\hat{p}^{1}\left(\left(q;\widehat{\mathbf{y}}\right)\right)\\
+\hat{p}^{k+1}\left(\left(q;\widehat{\mathbf{y}}\right)\right)
\end{array}\right)\widehat{\mathbf{g}}(\widehat{\mathbf{y}})\right]\\
 & + & {\displaystyle \sum_{k=2}^{N-1}}\left[{\int\limits _{\left[\underline{x},\bar{x}\right]^{N-1}:\widehat{y}_{k}\geq q>\widehat{y}_{k+1},\widehat{y}_{2}\geq x}}\left(\hat{p}^{k+1}\left(\left(q;\widehat{\mathbf{y}}\right)\right)\right)\widehat{\mathbf{g}}(\widehat{\mathbf{y}})\right].
\end{array}
\]
If $x<r$, then
\[
\Pi_{2}(q|x)={\int\limits _{\left[\underline{x},\bar{x}\right]^{N-1}:q>\widehat{y}_{1}}}\left(\hat{p}^{1}\left(\left(q;\widehat{\mathbf{y}}\right)\right)\right)\widehat{\mathbf{g}}(\widehat{\mathbf{y}})+{\displaystyle \sum_{k=1}^{N-1}}\left[{\int\limits _{\left[\underline{x},\bar{x}\right]^{N-1}:\widehat{y}_{k}\geq q>\widehat{y}_{k+1}}}\left(\hat{p}^{k+1}\left(\left(q;\widehat{\mathbf{y}}\right)\right)\right)\widehat{\mathbf{g}}(\widehat{\mathbf{y}})\right].
\]

\subsection{Convexity}

We show that the payoff $\Pi(q|x)$ is convex in its second argument
(the buyer's true type). The intuition is as follows: the derivative
of $\Pi(q|x)$ with respect to the buyer's type corresponds to the
probability that the buyer gets an item (either the first or the second).
Conditional on the report, that probability is increasing in the buyer's
type because a buyer with a higher valuation is more likely to win
the second auction if he does not win the first item.

Formally, the second derivative of the payoff $\Pi(q|x)$ with respect
to the buyer's true type, $\Pi_{22}(q|x)$, when $x\geq r$ and $q\geq x$
is given by
\[
\begin{array}{ccc}
\Pi_{22}(q|x) & = & {\int\limits _{\left[\underline{x},\bar{x}\right]^{N-1}:\widehat{y}_{1}=x}}\left(1-\hat{p}^{1}\left(\left(q;\widehat{\mathbf{y}}\right)\right)-\hat{p}^{2}\left(\left(q;\widehat{\mathbf{y}}\right)\right)\right)\widehat{\mathbf{g}}(\widehat{\mathbf{y}})\\
 & + & {\int\limits _{\left[\underline{x},\bar{x}\right]^{N-1}:q>\widehat{y}_{1},\widehat{y}_{2}=x}}\left(\hat{p}^{2}\left(\left(q;\widehat{\mathbf{y}}\right)\right)\right)\widehat{\mathbf{g}}(\widehat{\mathbf{y}})+{\int\limits _{\left[\underline{x},\bar{x}\right]^{N-1}:\widehat{y}_{1}\geq q,\widehat{y}_{2}=x}}\left(\hat{p}^{1}\left(\left(q;\widehat{\mathbf{y}}\right)\right)\right)\widehat{\mathbf{g}}(\widehat{\mathbf{y}}).
\end{array}
\]
The first integral represents the increase in the chance of getting
an item when the buyer's type moves from just below the highest competitor's
type $\widehat{y}_{1}$ to just above it: if $x>\widehat{y}_{1}$,
then the buyer gets an item for sure because he would win the second
auction. If $x<\widehat{y}_{1}$, then he gets an item only if he
or the highest competitor gets the first item. Similarly, the second
and third integrals represent the increase in the chance of getting
an item when the buyer's type moves from just below the second highest
competitor's type $\widehat{y}_{2}$ to just above it. Each of the
three integrals is weakly positive, so $\Pi_{22}(q|x)\geq0$.

Analogously, when $x\geq r$ and $q<x$, $\Pi_{22}(q|x)$ is given
by
\[
\begin{array}{ccc}
\Pi_{22}(q|x) & = & {\displaystyle \sum_{k=1}^{N-1}}\left[{\int\limits _{\left[\underline{x},\bar{x}\right]^{N-1}:\widehat{y}_{k}\geq q>\widehat{y}_{k+1},\widehat{y}_{1}=x}}\left(1-\hat{p}^{1}\left(\left(q;\widehat{\mathbf{y}}\right)\right)-\hat{p}^{k+1}\left(\left(q;\widehat{\mathbf{y}}\right)\right)\right)\widehat{\mathbf{g}}(\widehat{\mathbf{y}})\right]\\
 & + & {\displaystyle \sum_{k=2}^{N-1}}\left[{\int\limits _{\left[\underline{x},\bar{x}\right]^{N-1}:\widehat{y}_{k}\geq q>\widehat{y}_{k+1},\widehat{y}_{2}=x}}\left(\hat{p}^{1}\left(\left(q;\widehat{\mathbf{y}}\right)\right)\right)\widehat{\mathbf{g}}(\widehat{\mathbf{y}})\right]\geq0.
\end{array}
\]

Finally, if $x<r$, then $\Pi_{22}(q|x)=0$: the buyer will never
win the second auction, and his chance of getting the first item depends
on his report but not his true type. Thus, $\Pi(q|x)$ is convex in
the buyer's valuation, as desired.

\subsection{Incentive compatibility\label{subsec:Incentive-compatibility}}

Truthful reporting is a best response if and only if for all $x,q\in\left[\underline{x},\bar{x}\right]$,
\begin{equation}
U(x)=\Pi(x|x)-t(x)\geq(q|x)-t(q)=U(q)+\Pi(q|x)-\Pi(q|q)=U(q)+\intop_{q}^{x}\Pi_{2}(q|x^{\prime})dx^{\prime}.\label{eq:IC def}
\end{equation}

By substituting (\ref{eq:IC rewritten}) into (\ref{eq:IC def}),
we can rewrite the incentive compatibility condition as
\[
\int_{q}^{x}\Pi_{2}(x^{\prime}|x^{\prime})dx^{\prime}\geq\int_{q}^{x}\Pi_{2}(q|x^{\prime})dx^{\prime}.
\]
That condition holds if for any type $x$ and any reports $q,q'$
such that $q>x>q'$, we have $\Pi_{2}(q|x)\geq\Pi_{2}(x|x)\geq\Pi_{2}(q'|x)$.
The allocation rules in Theorems \ref{r above 1/2}, \ref{Theorem: optimal mechanism low r},
and \ref{Theorem: optimal mechanism high r} have the property that
$\hat{p}^{k}\left(\widehat{\mathbf{x}}\right)=0$ when $k>2$ for
all $\widehat{\mathbf{x}}$, so the expressions for $\Pi_{2}(q|x)-\Pi_{2}(x|x)$
simplify. If $x\geq r$ and $q>x$, then
\begin{equation}
\begin{array}{cc}
 & \Pi_{2}(q|x)-\Pi_{2}(x|x)\\
= & {\int\limits _{\left[\underline{x},\bar{x}\right]^{N-1}:\widehat{y}_{1}\geq x>\widehat{y}_{2}}}\left(\hat{p}^{1}\left(\left(q;\widehat{\mathbf{y}}\right)\right)+\hat{p}^{2}\left(\left(q;\widehat{\mathbf{y}}\right)\right)-\hat{p}^{1}\left(\left(x;\widehat{\mathbf{y}}\right)\right)-\hat{p}^{2}\left(\left(x;\widehat{\mathbf{y}}\right)\right)\right)\widehat{\mathbf{g}}(\widehat{\mathbf{y}})\\
+ & {\int\limits _{\left[\underline{x},\bar{x}\right]^{N-1}:q>\widehat{y}_{1}>\widehat{y}_{2}\geq x}}\left(\hat{p}^{1}\left(\left(q;\widehat{\mathbf{y}}\right)\right)\right)\widehat{\mathbf{g}}(\widehat{\mathbf{y}})+{\int\limits _{\left[\underline{x},\bar{x}\right]^{N-1}:\widehat{y}_{1}\geq q>\widehat{y}_{2}\geq x}}\left(\hat{p}^{2}\left(\left(q;\widehat{\mathbf{y}}\right)\right)\right)\widehat{\mathbf{g}}(\widehat{\mathbf{y}}).
\end{array}\label{eq:diff q>x}
\end{equation}
If $x\geq r$ and $q<x$, then
\begin{equation}
\begin{array}{cc}
 & \Pi_{2}(x|x)-\Pi_{2}(q|x)\\
= & {\int\limits _{\left[\underline{x},\bar{x}\right]^{N-1}:\widehat{y}_{1}\geq x>q>\widehat{y}_{2}}}\left(\hat{p}^{1}\left(\left(x;\widehat{\mathbf{y}}\right)\right)+\hat{p}^{2}\left(\left(x;\widehat{\mathbf{y}}\right)\right)-\hat{p}^{1}\left(\left(q;\widehat{\mathbf{y}}\right)\right)-\hat{p}^{2}\left(\left(q;\widehat{\mathbf{y}}\right)\right)\right)\widehat{\mathbf{g}}(\widehat{\mathbf{y}})\\
+ & {\int\limits _{\left[\underline{x},\bar{x}\right]^{N-1}:\widehat{y}_{1}\geq x>\widehat{y}_{2}\geq q}}\left(\hat{p}^{1}\left(\left(x;\widehat{\mathbf{y}}\right)\right)+\hat{p}^{2}\left(\left(x;\widehat{\mathbf{y}}\right)\right)-\hat{p}^{1}\left(\left(q;\widehat{\mathbf{y}}\right)\right)\right)\widehat{\mathbf{g}}(\widehat{\mathbf{y}}).
\end{array}\label{eq:diff q<x}
\end{equation}
Because the allocation rules in Theorems \ref{r above 1/2}, \ref{Theorem: optimal mechanism low r},
and \ref{Theorem: optimal mechanism high r} have the property that
$\hat{p}^{1}\left(\widehat{\mathbf{x}}\right)+\hat{p}^{2}\left(\widehat{\mathbf{x}}\right)$
is weakly increasing in $\widehat{x}_{1}$ and $\widehat{x}_{2}$,
(\ref{eq:diff q>x}) and (\ref{eq:diff q<x}) are positive, as desired.

Next consider the case $x<r$. The allocation rules in Theorems \ref{r above 1/2},
\ref{Theorem: optimal mechanism low r}, and \ref{Theorem: optimal mechanism high r}
have the additional property that $\hat{p}^{2}\left(\widehat{\mathbf{x}}\right)=0$
when $\widehat{x}_{2}<r$, so for $q>x$ we have
\begin{equation}
\begin{array}{cc}
 & \Pi_{2}(q|x)-\Pi_{2}(x|x)\\
= & {\int\limits _{\left[\underline{x},\bar{x}\right]^{N-1}:x>\widehat{y}_{1}}}\left(\hat{p}^{1}\left(\left(q;\widehat{\mathbf{y}}\right)\right)-\hat{p}^{1}\left(\left(x;\widehat{\mathbf{y}}\right)\right)\right)\widehat{\mathbf{g}}(\widehat{\mathbf{y}})\\
+ & {\int\limits _{\left[\underline{x},\bar{x}\right]^{N-1}:q>\widehat{y}_{1}\geq x}}\left(\hat{p}^{1}\left(\left(q;\widehat{\mathbf{y}}\right)\right)\right)\widehat{\mathbf{g}}(\widehat{\mathbf{y}})+{\int\limits _{\left[\underline{x},\bar{x}\right]^{N-1}:\widehat{y}_{1}\geq q}}\left(\hat{p}^{2}\left(\left(q;\widehat{\mathbf{y}}\right)\right)\right)\widehat{\mathbf{g}}(\widehat{\mathbf{y}}).
\end{array}\label{eq:diff x<r q>x}
\end{equation}

Finally, if $x<r$ and $q<x$, then
\begin{equation}
\begin{array}{cc}
 & \Pi_{2}(x|x)-\Pi_{2}(q|x)\\
= & {\int\limits _{\left[\underline{x},\bar{x}\right]^{N-1}:q>\widehat{y}_{1}}}\left(\hat{p}^{1}\left(\left(x;\widehat{\mathbf{y}}\right)\right)-\hat{p}^{1}\left(\left(q;\widehat{\mathbf{y}}\right)\right)\right)\widehat{\mathbf{g}}(\widehat{\mathbf{y}})+{\int\limits _{\left[\underline{x},\bar{x}\right]^{N-1}:x>\widehat{y}_{1}\geq q}}\left(\hat{p}^{1}\left(\left(x;\widehat{\mathbf{y}}\right)\right)\right)\widehat{\mathbf{g}}(\widehat{\mathbf{y}}).
\end{array}\label{eq:diff r>x>q}
\end{equation}
Because the specified allocation rules have the property that $\hat{p}^{1}\left(\widehat{\mathbf{x}}\right)$
is weakly increasing in $\widehat{x}_{1}$, (\ref{eq:diff x<r q>x})
and (\ref{eq:diff r>x>q}) are both positive as well. We conclude
that the mechanisms in Theorems \ref{r above 1/2}, \ref{Theorem: optimal mechanism low r},
and \ref{Theorem: optimal mechanism high r} are incentive compatible.
It remains only to show that the allocation rules solve the seller's
revenue maximization problem. We made that argument for the $r\leq\underline{x}$
case in Section \ref{sec:Optimal Mechanism} and for the $r\in\left[\psi^{-1}(0),\bar{x}\right)$
case in Section \ref{subsec:If r>psi-1(0)}. We cover the other cases
next.

\subsection{Constrained Optimization when $r\in\left(\underline{x},\psi^{-1}(0)\right)$\label{sec:Proving-Theorems-r>0}}

Recall that the seller's problem is to maximize (\ref{eqrevenue-1})
subject to $\int_{q}^{r}\left[\Pi_{2}(x^{\prime}|x^{\prime})-\Pi_{2}(q|x^{\prime})\right]dx^{\prime}\geq0$
for each $q\in\left[\underline{x},r\right)$ (with Lagrange multiplier
$\lambda_{r,q}$), and to
\[
\Pi_{2}(x|x)-\limsup_{\epsilon\searrow0}\Pi_{2}(x-\epsilon|x)\geq0
\]
for each $x\in\left(\underline{x},a(r)\right]$ (multiplier $\mu_{x}$).
Using the derivations in Section \ref{subsec:Pi-2}, we write out
\[
\begin{array}{c}
\Pi_{2}(x|x)-\limsup_{\epsilon\searrow0}\Pi_{2}(x-\epsilon|x)=\\
\limsup_{\epsilon\searrow0}\left\{ \begin{array}{cc}
 & {\int\limits _{\left[\underline{x},\bar{x}\right]^{N-1}:x>\widehat{y}_{1}}}\left[\hat{p}^{1}\left(\left(x;\widehat{\mathbf{y}}\right)\right)-\hat{p}^{1}\left(\left(x-\epsilon;\widehat{\mathbf{y}}\right)\right)\right]\widehat{\mathbf{g}}(\widehat{\mathbf{y}})\\
+ & \sum_{k=1}^{N-1}\left[\int_{\left[\underline{x},\bar{x}\right]^{N-1}:\widehat{y}_{k}\geq x>\widehat{y}_{k+1}}\left[\hat{p}^{k+1}\left(\left(x;\widehat{\mathbf{y}}\right)\right)-\hat{p}^{k+1}\left(\left(x-\epsilon;\widehat{\mathbf{y}}\right)\right)\right]\widehat{\mathbf{g}}(\widehat{\mathbf{y}})\right]
\end{array}\right\} 
\end{array}
\]
for each $x\in\left[\underline{x},r\right)$;
\[
\begin{array}{c}
\Pi_{2}(x|x)-\limsup_{\epsilon\searrow0}\Pi_{2}(x-\epsilon|x)=\\
\limsup_{\epsilon\searrow0}\left\{ \begin{array}{cc}
 & {\int\limits _{\left[\underline{x},\bar{x}\right]^{N-1}:\widehat{y}_{1}\geq x>\widehat{y}_{2}}}\left[\begin{array}{c}
\left(\hat{p}^{1}\left(\left(x;\widehat{\mathbf{y}}\right)\right)-\hat{p}^{1}\left(\left(x-\epsilon;\widehat{\mathbf{y}}\right)\right)\right)\\
+\left(\hat{p}^{2}\left(\left(x;\widehat{\mathbf{y}}\right)\right)-\hat{p}^{2}\left(\left(x-\epsilon;\widehat{\mathbf{y}}\right)\right)\right)
\end{array}\right]\widehat{\mathbf{g}}(\widehat{\mathbf{y}})\\
+ & \sum_{k=2}^{N-1}\left[\int_{\left[\underline{x},\bar{x}\right]^{N-1}:\widehat{y}_{k}\geq x>\widehat{y}_{k+1}}\left(\hat{p}^{k+1}\left(\left(x;\widehat{\mathbf{y}}\right)\right)-\hat{p}^{k+1}\left(\left(x-\epsilon;\widehat{\mathbf{y}}\right)\right)\right)\widehat{\mathbf{g}}(\widehat{\mathbf{y}})\right]
\end{array}\right\} 
\end{array}
\]
for each $x\in\left[r,a(r)\right]$;
\[
\begin{array}{c}
\Pi_{2}(r|r)-\Pi_{2}(q|r)=\\
\begin{array}{cc}
 & \begin{array}{cc}
 & \int_{\left[\underline{x},\bar{x}\right]^{N-1}:\widehat{y}_{1}\geq r>\widehat{y}_{2}}\left(\hat{p}^{1}\left(\left(r;\widehat{\mathbf{y}}\right)\right)+\hat{p}^{2}\left(\left(r;\widehat{\mathbf{y}}\right)\right)\right)\widehat{\mathbf{g}}(\widehat{\mathbf{y}})\\
+ & \sum_{k=2}^{N-1}\left[\int_{\left[\underline{x},\bar{x}\right]^{N-1}:\widehat{y}_{k}\geq r>\widehat{y}_{k+1}}\hat{p}^{k+1}\left(\left(r;\widehat{\mathbf{y}}\right)\right)\widehat{\mathbf{g}}(\widehat{\mathbf{y}})\right]
\end{array}\\
- & \left[\begin{array}{cc}
 & \int_{\left[\underline{x},\bar{x}\right]^{N-1}:\widehat{y}_{1}\geq r,\widehat{y}_{2}<q}\left(\hat{p}^{1}\left(\left(q;\widehat{\mathbf{y}}\right)\right)+\hat{p}^{2}\left(\left(q;\widehat{\mathbf{y}}\right)\right)\right)\widehat{\mathbf{g}}(\widehat{\mathbf{y}})\\
+ & \sum_{k=1}^{N-1}\left[\int_{\left[\underline{x},\bar{x}\right]^{N-1}:\widehat{y}_{1}\geq r,\widehat{y}_{2}<r,\widehat{y}_{k}\geq q>\widehat{y}_{k+1}}\left(\begin{array}{c}
\hat{p}^{1}\left(\left(q;\widehat{\mathbf{y}}\right)\right)\\
+\hat{p}^{k+1}\left(\left(q;\widehat{\mathbf{y}}\right)\right)
\end{array}\right)\widehat{\mathbf{g}}(\widehat{\mathbf{y}})\right]\\
+ & \sum_{k=2}^{N-1}\left[\int_{\left[\underline{x},\bar{x}\right]^{N-1}:\widehat{y}_{2}\geq r,\widehat{y}_{k}\geq q>\widehat{y}_{k+1}}\hat{p}^{k+1}\left(\left(q;\widehat{\mathbf{y}}\right)\right)\widehat{\mathbf{g}}(\widehat{\mathbf{y}})\right]
\end{array}\right]
\end{array}
\end{array}
\]
for each $q\in\left[\underline{x},r\right)$; and
\[
\begin{array}{c}
\Pi_{2}(x|x)-\Pi_{2}(q|x)=\\
\begin{array}{cc}
 & \begin{array}{cc}
 & {\int\limits _{\left[\underline{x},\bar{x}\right]^{N-1}:x>\widehat{y}_{1}}}\left(\hat{p}^{1}\left(\left(x;\widehat{\mathbf{y}}\right)\right)\right)\widehat{\mathbf{g}}(\widehat{\mathbf{y}})\\
+ & {\displaystyle \sum_{k=1}^{N-1}}\left[{\int\limits _{\left[\underline{x},\bar{x}\right]^{N-1}:\widehat{y}_{k}\geq x>\widehat{y}_{k+1}}}\hat{p}^{k+1}\left(\left(x;\widehat{\mathbf{y}}\right)\right)\widehat{\mathbf{g}}(\widehat{\mathbf{y}})\right]
\end{array}\\
- & \left[\begin{array}{cc}
 & {\int\limits _{\left[\underline{x},\bar{x}\right]^{N-1}:q>\widehat{y}_{1}}}\left(\hat{p}^{1}\left(\left(q;\widehat{\mathbf{y}}\right)\right)\right)\widehat{\mathbf{g}}(\widehat{\mathbf{y}})\\
+ & {\displaystyle \sum_{k=1}^{N-1}}\left[{\int\limits _{\left[\underline{x},\bar{x}\right]^{N-1}:\widehat{y}_{k}\geq q>\widehat{y}_{k+1}}}\hat{p}^{k+1}\left(\left(q;\widehat{\mathbf{y}}\right)\right)\widehat{\mathbf{g}}(\widehat{\mathbf{y}})\right]
\end{array}\right]
\end{array}
\end{array}
\]
for each $x\in\left[\underline{x},r\right)$ and $q\in\left[\underline{x},x\right)$.
Note that for $x\geq r$, the integrals do not include the case $x>\widehat{y}_{1}$,
because the highest-type buyer is sure to get an object when his type
is above $r$.

\subsubsection{When $Z^{r}(r)\protect\geq0$}

In this case, Theorem \ref{Theorem: optimal mechanism high r} specifies
allocation whenever either i) $\widehat{x}_{1}\geq r>\widehat{x}_{3}$,
or ii) $\widehat{x}_{3}\geq r$ and $\psi\left(\widehat{x}_{2}\right)+\widehat{x}_{2}-\widehat{x}_{3}\geq0$,
and not otherwise. We make the following guesses for the values of
the Lagrange multipliers: for all $x\in\left[\underline{x},r\right)$,
$\lambda_{r,x}=\mu_{x}=0;$ and for $x\in\left[r,a(r)\right]$, $\begin{array}{c}
\mu_{x}=\int_{x}^{a(r)}N\cdot\left[r-x'-\psi(x')\right]f(x^{\prime})dx^{\prime}.\end{array}$ That is, only the immediate downward constraints for type $r$ and
above bind. The intuition behind that guess for the values of $\mu_{x}$
is as follows: suppose that we relaxed the constraint that type $x\in\left[r,a(r)\right]$
must have a weakly higher chance of getting an object if he reports
truthfully than if he underreports to $x-\epsilon$. Then the seller
could not allocate when $\widehat{x}_{2}=x$ and $\widehat{x}_{3}<r$,
and thus avoid earning the negative marginal revenue $x+\psi(x)-r$
in that case. There is an additional benefit: for type $x+\epsilon$,
now underreporting does not lead to allocation, and so the seller
is free to not allocate when $\widehat{x}_{2}=x+\epsilon$ and $\widehat{x}_{3}<r$,
without violating the constraint for type $x+\epsilon$. Iterating,
we see that relaxing the constraint for the \emph{single} type $x$
allows the seller to avoid the negative marginal revenue $x'+\psi(x')-r$
for \emph{every} type $x'$ between $x$ and $a(r)$.

In what follows, the key feature of $\mu_{x}$ is that $\mu_{x}-\mu_{x+0}=N\cdot\left[r-x-\psi(x)\right]f(x)dx.$
For example, suppose that $\widehat{x}_{2}\in\left[r,a(r)\right)$.
Allocating to either of the top two bidders in that case helps with
the constraint for type $\widehat{x}_{2}$ (he gets an item by telling
the truth), but it hurts with the constraint for a slightly higher
type (he could get an item by underreporting his type as $\widehat{x}_{2}$).
The net marginal effect is the difference between $\mu_{\widehat{x}_{2}}$
and $\mu_{\widehat{x}_{2}+0}$.

We use that key feature repeatedly as we next take the partial derivative
of the seller's expected revenue with respect to $\hat{p}^{k}\left(\hat{\mathbf{x}}\right)$,
given any vector of ordered types $\hat{\mathbf{x}}$, and plug in
those guesses. Note that for any $\widehat{\mathbf{x}}$ and any $k$,
$N\cdot f(\widehat{x}_{k})\cdot\hat{\mathbf{g}}(\widehat{\mathbf{x}}_{-k})=\hat{\mathbf{f}}(\widehat{\mathbf{x}})$.
\begin{enumerate}
\item If $\widehat{x}_{2}\geq a(r)$, then $\frac{\partial ER(\widehat{X})}{\delta\hat{p}^{1}\left(\hat{\mathbf{x}}\right)}=\frac{\partial ER(\widehat{X})}{\delta\hat{p}^{2}\left(\hat{\mathbf{x}}\right)}=\left[\widehat{x}_{2}+\psi\left(\widehat{x}_{2}\right)-\max\left\{ \widehat{x}_{3},r\right\} \right]\hat{\mathbf{f}}(\widehat{\mathbf{x}}).$
\item If $\widehat{x}_{2}\in\left[r,a(r)\right)$, then
\begin{align*}
\frac{\partial ER(\widehat{X})}{\delta\hat{p}^{1}\left(\hat{\mathbf{x}}\right)}=\frac{\partial ER(\widehat{X})}{\delta\hat{p}^{2}\left(\hat{\mathbf{x}}\right)} & =\left[\widehat{x}_{2}+\psi\left(\widehat{x}_{2}\right)-\max\left\{ \widehat{x}_{3},r\right\} \right]\hat{\mathbf{f}}(\widehat{\mathbf{x}})+\hat{\mathbf{g}}(\widehat{\mathbf{x}}_{-2})\mu_{\widehat{x}_{2}}-\hat{\mathbf{g}}(\widehat{\mathbf{x}}_{-2})\mu_{\widehat{x}_{2}+0}\\
 & =\left[\widehat{x}_{2}+\psi\left(\widehat{x}_{2}\right)-\max\left\{ \widehat{x}_{3},r\right\} \right]\hat{\mathbf{f}}(\widehat{\mathbf{x}})\\
 & +\hat{\mathbf{g}}(\widehat{\mathbf{x}}_{-2})N\cdot\left[r-\widehat{x}_{2}-\psi(\widehat{x}_{2})\right]f(\widehat{x}_{2})\\
 & =\hat{\mathbf{f}}(\widehat{\mathbf{x}})\cdot\left[\widehat{x}_{2}+\psi\left(\widehat{x}_{2}\right)-\max\left\{ \widehat{x}_{3},r\right\} +\left(r-\widehat{x}_{2}-\psi(\widehat{x}_{2})\right)\right]\\
 & =\hat{\mathbf{f}}(\widehat{\mathbf{x}})\cdot\left[r-\max\left\{ \widehat{x}_{3},r\right\} \right].
\end{align*}
\item \label{enu:x1>a, x2<r}If $\widehat{x}_{1}\geq r$ and $\widehat{x}_{2}<r$,
then $\frac{\partial ER(\widehat{X})}{\delta\hat{p}^{1}\left(\hat{\mathbf{x}}\right)}=r\hat{\mathbf{f}}(\widehat{\mathbf{x}})>0.$
\item In every case above, for all $k>2$ such that $\widehat{x}_{k}\geq a(r)$,
\[
\frac{\partial ER(\widehat{X})}{\delta\hat{p}^{k}\left(\hat{\mathbf{x}}\right)}=\psi\left(\widehat{x}_{k}\right)\hat{\mathbf{f}}(\widehat{\mathbf{x}})\leq\frac{\partial ER(\widehat{X})}{\delta\hat{p}^{2}\left(\hat{\mathbf{x}}\right)};
\]
for all $k>2$ such that $\widehat{x}_{k}\in\left[r,a(r)\right)$,
\begin{align*}
\frac{\partial ER(\widehat{X})}{\delta\hat{p}^{k}\left(\hat{\mathbf{x}}\right)} & =\psi\left(\widehat{x}_{k}\right)\hat{\mathbf{f}}(\widehat{\mathbf{x}})+\hat{\mathbf{g}}(\widehat{\mathbf{x}}_{-k})\mu_{\widehat{x}_{k}}-\hat{\mathbf{g}}(\widehat{\mathbf{x}}_{-k})\mu_{\widehat{x}_{k}+0}\\
 & =\psi\left(\widehat{x}_{k}\right)\hat{\mathbf{f}}(\widehat{\mathbf{x}})+\hat{\mathbf{g}}(\widehat{\mathbf{x}}_{-k})N\cdot\left[r-\widehat{x}_{k}-\psi(\widehat{x}_{k})\right]f(\widehat{x}_{k})\\
 & =\hat{\mathbf{f}}(\widehat{\mathbf{x}})\cdot\left[\psi\left(\widehat{x}_{k}\right)+\left(r-\widehat{x}_{k}-\psi(\widehat{x}_{k})\right)\right]=\hat{\mathbf{f}}(\widehat{\mathbf{x}})\cdot\left[r-\widehat{x}_{k}\right]<0;
\end{align*}
and for all $k$ such that $\widehat{x}_{k}<r$, $\frac{\partial ER(\widehat{X})}{\delta\hat{p}^{k}\left(\hat{\mathbf{x}}\right)}=\psi\left(\widehat{x}_{k}\right)\hat{\mathbf{f}}(\widehat{\mathbf{x}})<0.$
\end{enumerate}
The marginal revenues above are weakly positive in each case where
Theorem \ref{Theorem: optimal mechanism high r} specifies allocation,
and they are weakly negative in each case where Theorem \ref{Theorem: optimal mechanism high r}
specifies no allocation. Thus, our guesses for the values of the Lagrange
multipliers, together with the allocation rule in Theorem \ref{Theorem: optimal mechanism high r},
form a solution to the seller's constrained optimization problem.

\subsubsection{When $Z^{r}(r)\protect\leq0$}

In this case, Theorem \ref{Theorem: optimal mechanism low r} specifies
allocation if and only if $\psi\left(\widehat{x}_{2}\right)+\widehat{x}_{2}-\max\left\{ r,\widehat{x}_{3}\right\} \geq0$.
The derivative of $Z^{r}(x^{*})$ is given by $z^{r}\left(x^{*}\right)f\left(x^{*}\right)$,
where the function $z^{r}\left(x^{*}\right)$ is defined as
\[
z^{r}(x^{*})\equiv-rF\left(r\right)+(N-1)\int\limits _{r}^{\min\left\{ x^{*},a(r)\right\} }[r-x'-\psi(x')]f(x')dx'.
\]
The function $z^{r}\left(x^{*}\right)$ is strictly increasing for
$x^{*}<a(r)$ and constant for $x^{*}\geq a(r)$. When $Z^{r}\left(\bar{x}\right)-Z^{r}(r)=0-Z^{r}(r)\geq0$,
therefore, it must be that $z^{r}\left(a(r)\right)\geq0$: otherwise
$Z^{r}(x^{*})$ would be strictly decreasing throughout. We will use
the inequality $z^{r}\left(a(r)\right)\geq0$ below.

We make the following guesses for the Lagrange multipliers: for $x\in\left[\underline{x},r\right)$,
$\begin{array}{c}
\lambda_{r,x}=r\frac{N}{N-1}f\left(x\right)\end{array}$ and $\begin{array}{c}
\mu_{x}=\int_{\underline{x}}^{x}r\frac{N}{N-1}F(x^{\prime})dx^{\prime};\end{array}$and for $x\in\left(r,a(r)\right]$, $\begin{array}{c}
\mu_{x}=\int_{x}^{a(r)}N\cdot\left[r-x'-\psi(x')\right]f(x^{\prime})dx^{\prime}.\end{array}$ The differences relative to the $Z^{r}(r)\geq0$ case are that now
all the downward constraints bind for type $r$, and the immediate
downward constraints bind for types below $r$. Allocating to either
of the top two bidders when $\widehat{x}_{2}=r$ helps with all the
downward constraints for type $r$, because he gets an item by telling
the truth. On the other hand, allocating to the highest bidder when
$\widehat{x}_{1}\geq r>\widehat{x}_{2}$ hurts with a constraint,
because a bidder with type $r$ gets an object by underreporting his
type as $\widehat{x}_{2}$. The intuition for our guess of the value
of $\lambda_{r,x}$ is that if we relaxed the constraint, then the
seller could allocate to the high bidder whenever $\widehat{x}_{1}\geq r>\widehat{x}_{2}$
and earn the corresponding marginal revenue $r$.

As we take the partial derivative of the seller's expected revenue
with respect to $\hat{p}^{k}\left(\hat{\mathbf{x}}\right)$ and plug
in those guesses, we again use the feature that $\mu_{x}-\mu_{x+0}=N\cdot\left[r-x-\psi(x)\right]f(x)dx$
for $x\in\left(r,a(r)\right]$. Similarly, we use the feature that
for $x\in\left[\underline{x},r\right)$, $\mu_{x+0}-\mu_{x}=r\frac{N}{N-1}F(x)dx.$
\begin{enumerate}
\item If $\widehat{x}_{2}\geq a(r)$, then
\[
\frac{\partial ER(\widehat{X})}{\delta\hat{p}^{1}\left(\hat{\mathbf{x}}\right)}=\left[\widehat{x}_{2}+\psi\left(\widehat{x}_{2}\right)-\max\left\{ \widehat{x}_{3},r\right\} \right]\hat{\mathbf{f}}(\widehat{\mathbf{x}})-\sum_{k:\widehat{x}_{k}<r}\left[\hat{\mathbf{g}}(\widehat{\mathbf{x}}_{-k})\lambda_{r,\widehat{x}_{k}}\right]\leq\frac{\partial ER(\widehat{X})}{\delta\hat{p}^{2}\left(\hat{\mathbf{x}}\right)};
\]
\[
\frac{\partial ER(\widehat{X})}{\delta\hat{p}^{2}\left(\hat{\mathbf{x}}\right)}=\left[\widehat{x}_{2}+\psi\left(\widehat{x}_{2}\right)-\max\left\{ \widehat{x}_{3},r\right\} \right]\hat{\mathbf{f}}(\widehat{\mathbf{x}}).
\]
\item If $\widehat{x}_{2}\in\left(r,a(r)\right)$, then
\[
\frac{\partial ER(\widehat{X})}{\delta\hat{p}^{1}\left(\hat{\mathbf{x}}\right)}=\left(\begin{array}{c}
\left[\widehat{x}_{2}+\psi\left(\widehat{x}_{2}\right)-\max\left\{ \widehat{x}_{3},r\right\} \right]\hat{\mathbf{f}}(\widehat{\mathbf{x}})+\hat{\mathbf{g}}(\widehat{\mathbf{x}}_{-2})\mu_{\widehat{x}_{2}}\\
-\hat{\mathbf{g}}(\widehat{\mathbf{x}}_{-2})\mu_{\widehat{x}_{2}+0}-\sum_{k:\widehat{x}_{k}<r}\left[\hat{\mathbf{g}}(\widehat{\mathbf{x}}_{-k})\lambda_{r,\widehat{x}_{k}}\right]
\end{array}\right)\leq\frac{\partial ER(\widehat{X})}{\delta\hat{p}^{2}\left(\hat{\mathbf{x}}\right)};
\]
\begin{align*}
\frac{\partial ER(\widehat{X})}{\delta\hat{p}^{2}\left(\hat{\mathbf{x}}\right)} & =\left[\widehat{x}_{2}+\psi\left(\widehat{x}_{2}\right)-\max\left\{ \widehat{x}_{3},r\right\} \right]\hat{\mathbf{f}}(\widehat{\mathbf{x}})+\hat{\mathbf{g}}(\widehat{\mathbf{x}}_{-2})\mu_{\widehat{x}_{2}}-\hat{\mathbf{g}}(\widehat{\mathbf{x}}_{-2})\mu_{\widehat{x}_{2}+0}\\
 & =\left[\widehat{x}_{2}+\psi\left(\widehat{x}_{2}\right)-\max\left\{ \widehat{x}_{3},r\right\} \right]\hat{\mathbf{f}}(\widehat{\mathbf{x}})+\hat{\mathbf{g}}(\widehat{\mathbf{x}}_{-2})N\cdot\left[r-\widehat{x}_{2}-\psi(\widehat{x}_{2})\right]f(\widehat{x}_{2})\\
 & =\hat{\mathbf{f}}(\widehat{\mathbf{x}})\cdot\left[\widehat{x}_{2}+\psi\left(\widehat{x}_{2}\right)-\max\left\{ \widehat{x}_{3},r\right\} +\left(r-\widehat{x}_{2}-\psi(\widehat{x}_{2})\right)\right]\\
 & =\hat{\mathbf{f}}(\widehat{\mathbf{x}})\cdot\left[r-\max\left\{ \widehat{x}_{3},r\right\} \right].
\end{align*}
\item If $\widehat{x}_{2}=r$, then
\begin{align*}
\frac{\partial ER(\widehat{X})}{\delta\hat{p}^{1}\left(\hat{\mathbf{x}}\right)} & =\left[r+\psi\left(r\right)-r\right]\hat{\mathbf{f}}(\widehat{\mathbf{x}})-\hat{\mathbf{g}}(\widehat{\mathbf{x}}_{-2})\mu_{r+0}+\hat{\mathbf{g}}(\widehat{\mathbf{x}}_{-2})\int_{\underline{x}}^{r}\lambda_{r,x^{\prime}}dx^{\prime}-\sum_{k=3}^{N}\left[\hat{\mathbf{g}}(\widehat{\mathbf{x}}_{-k})\lambda_{r,\widehat{x}_{k}}\right]\\
 & \leq\frac{\partial ER(\widehat{X})}{\delta\hat{p}^{2}\left(\hat{\mathbf{x}}\right)};
\end{align*}
\begin{align*}
\frac{\partial ER(\widehat{X})}{\delta\hat{p}^{2}\left(\hat{\mathbf{x}}\right)} & =\left[r+\psi\left(r\right)-r\right]\hat{\mathbf{f}}(\widehat{\mathbf{x}})-\hat{\mathbf{g}}(\widehat{\mathbf{x}}_{-2})\mu_{r+0}+\hat{\mathbf{g}}(\widehat{\mathbf{x}}_{-2})\int_{\underline{x}}^{r}\lambda_{r,x^{\prime}}dx^{\prime}\\
 & =\left[r+\psi\left(r\right)-r\right]\hat{\mathbf{g}}(\widehat{\mathbf{x}}_{-2})Nf(r)-\hat{\mathbf{g}}(\widehat{\mathbf{x}}_{-2})\int_{r+0}^{a(r)}N\cdot\left[r-x-\psi(x)\right]f(x)dx\\
 & +\hat{\mathbf{g}}(\widehat{\mathbf{x}}_{-2})F(r)\frac{N}{N-1}r\\
 & =-\hat{\mathbf{g}}(\widehat{\mathbf{x}}_{-2})\int_{r}^{a(r)}N\cdot\left[r-x-\psi(x)\right]f(x)dx+\hat{\mathbf{g}}(\widehat{\mathbf{x}}_{-2})F(r)\frac{N}{N-1}r\\
 & =\hat{\mathbf{g}}(\widehat{\mathbf{x}}_{-2})\frac{N}{N-1}\left[F(r)r-(N-1)\int_{r}^{a(r)}\left[r-x-\psi(x)\right]f(x)dx\right]\\
 & =-\hat{\mathbf{g}}(\widehat{\mathbf{x}}_{-2})\frac{N}{N-1}z^{r}\left(a(r)\right)\leq0.
\end{align*}
\item If $\widehat{x}_{1}\geq r$ and $\widehat{x}_{2}<r$, then
\begin{align*}
\frac{\partial ER(\widehat{X})}{\delta\hat{p}^{1}\left(\hat{\mathbf{x}}\right)} & =r\hat{\mathbf{f}}(\widehat{\mathbf{x}})-\sum_{k=2}^{N}\left[\hat{\mathbf{g}}(\widehat{\mathbf{x}}_{-k})\lambda_{r,\widehat{x}_{k}}\right]=r\hat{\mathbf{f}}(\widehat{\mathbf{x}})-\sum_{k=2}^{N}\left[\hat{\mathbf{f}}(\widehat{\mathbf{x}})\frac{1}{N-1}r\right]=\hat{\mathbf{f}}(\widehat{\mathbf{x}})\cdot\left[r-r\right]=0.
\end{align*}
\item In every case above, for all $k>2$ such that $\widehat{x}_{k}\geq a(r)$,
\begin{align*}
\frac{\partial ER(\widehat{X})}{\delta\hat{p}^{k}\left(\hat{\mathbf{x}}\right)} & =\psi\left(\widehat{x}_{k}\right)\hat{\mathbf{f}}(\widehat{\mathbf{x}})\leq\frac{\partial ER(\widehat{X})}{\delta\hat{p}^{2}\left(\hat{\mathbf{x}}\right)};
\end{align*}
for all $k>2$ such that $\widehat{x}_{k}\in\left(r,a(r)\right]$,
\begin{align*}
\frac{\partial ER(\widehat{X})}{\delta\hat{p}^{k}\left(\hat{\mathbf{x}}\right)} & =\psi\left(\widehat{x}_{k}\right)\hat{\mathbf{f}}(\widehat{\mathbf{x}})+\hat{\mathbf{g}}(\widehat{\mathbf{x}}_{-k})\mu_{\widehat{x}_{k}}-\hat{\mathbf{g}}(\widehat{\mathbf{x}}_{-k})\mu_{\widehat{x}_{k}+0}\\
 & =\psi\left(\widehat{x}_{k}\right)\hat{\mathbf{f}}(\widehat{\mathbf{x}})+\hat{\mathbf{g}}(\widehat{\mathbf{x}}_{-k})N\cdot\left[r-\widehat{x}_{k}-\psi(\widehat{x}_{k})\right]f(\widehat{x}_{k})\\
 & =\hat{\mathbf{f}}(\widehat{\mathbf{x}})\cdot\left[\psi\left(\widehat{x}_{k}\right)+\left(r-\widehat{x}_{k}-\psi(\widehat{x}_{k})\right)\right]=\hat{\mathbf{f}}(\widehat{\mathbf{x}})\cdot\left[r-\widehat{x}_{k}\right]<0;
\end{align*}
for all $k>2$ such that $\widehat{x}_{k}=r$,
\begin{align*}
\frac{\partial ER(\widehat{X})}{\delta\hat{p}^{k}\left(\hat{\mathbf{x}}\right)} & =\left[\psi\left(r\right)\right]\hat{\mathbf{f}}(\widehat{\mathbf{x}})-\hat{\mathbf{g}}(\widehat{\mathbf{x}}_{-k})\mu_{r+0}+\hat{\mathbf{g}}(\widehat{\mathbf{x}}_{-k})\int_{\underline{x}}^{r}\lambda_{r,x^{\prime}}dx^{\prime}\\
 & =\left[r+\psi\left(r\right)-r\right]\hat{\mathbf{g}}(\widehat{\mathbf{x}}_{-k})Nf(r)-\hat{\mathbf{g}}(\widehat{\mathbf{x}}_{-k})\int_{r+0}^{a(r)}N\cdot\left[r-x-\psi(x)\right]f(x)dx\\
 & +\hat{\mathbf{g}}(\widehat{\mathbf{x}}_{-k})F(r)\frac{N}{N-1}r\\
 & =-\hat{\mathbf{g}}(\widehat{\mathbf{x}}_{-k})\int_{r}^{a(r)}N\cdot\left[r-x-\psi(x)\right]f(x)dx+\hat{\mathbf{g}}(\widehat{\mathbf{x}}_{-k})F(r)\frac{N}{N-1}r\\
 & =\hat{\mathbf{g}}(\widehat{\mathbf{x}}_{-k})\frac{N}{N-1}\left[F(r)r-(N-1)\int_{r}^{a(r)}N\cdot\left[r-x-\psi(x)\right]f(x)dx\right]\\
 & =-\hat{\mathbf{g}}(\widehat{\mathbf{x}}_{-k})\frac{N}{N-1}z^{r}\left(a(r)\right)\leq0;
\end{align*}
and for all $k$ such that $\widehat{x}_{k}<r$, 
\begin{align*}
\frac{\partial ER(\widehat{X})}{\delta\hat{p}^{k}\left(\hat{\mathbf{x}}\right)} & =\psi\left(\widehat{x}_{k}\right)\hat{\mathbf{f}}(\widehat{\mathbf{x}})-\hat{\mathbf{g}}(\widehat{\mathbf{x}}_{-k})\mu_{\widehat{x}_{k}+0}+\hat{\mathbf{g}}(\widehat{\mathbf{x}}_{-k})\mu_{\widehat{x}_{k}}\\
 & +\hat{\mathbf{g}}(\widehat{\mathbf{x}}_{-k})\int_{\underline{x}}^{\widehat{x}_{k}}\lambda_{r,x^{\prime}}dx^{\prime}-\hat{\mathbf{g}}(\widehat{\mathbf{x}}_{-k})\int_{\widehat{x}_{k}}^{r}\lambda_{r,\widehat{x}_{k}}dx^{\prime}\\
 & =\psi\left(\widehat{x}_{k}\right)\hat{\mathbf{f}}(\widehat{\mathbf{x}})-\hat{\mathbf{g}}(\widehat{\mathbf{x}}_{-k})\left[r\frac{N}{N-1}F(\widehat{x}_{k})\right]\\
 & +\hat{\mathbf{g}}(\widehat{\mathbf{x}}_{-k})F(\widehat{x}_{k})\frac{N}{N-1}r-\hat{\mathbf{g}}(\widehat{\mathbf{x}}_{-k})\int_{\widehat{x}_{k}}^{r}\lambda_{r,\widehat{x}_{k}}dx^{\prime}\leq\psi\left(\widehat{x}_{k}\right)\hat{\mathbf{f}}(\widehat{\mathbf{x}})<0.
\end{align*}
\end{enumerate}
The marginal revenues above are weakly positive in each case where
Theorem \ref{Theorem: optimal mechanism low r} specifies allocation,
and they are weakly negative in each case where Theorem \ref{Theorem: optimal mechanism low r}
specifies no allocation. Thus, our guesses for the values of the Lagrange
multipliers, together with the allocation rule in Theorem \ref{Theorem: optimal mechanism low r},
form a solution to the seller's constrained optimization problem.

\appendix
\begin{center}
{\Huge{}\pagebreak{}{}Online Appendix}{\Huge\par}
\par\end{center}

\section{Modified Pay-Your-Bid Auction\label{sec:Modified-Pay-Your-Bid-Auction}}

As the first step in constructing the equilibrium bid function $\beta$,
let $H(q)$ denote the probability that a buyer who submits a bid
of $\beta(q)$ in the auction will pay his bid, conditional on other
buyers bidding according to $\beta(\cdot)$. The bidder pays his bid
if $\beta(q)$ is the highest bid $(q>Y_{(1)})$ or if $\beta(q)$
is the second-highest bid and the good is allocated $(Y_{(1)}>q\geq a\left(Y_{(2)}\right))$.
Recalling that $G_{k}$ is the distribution of $Y_{(k)}$, the $k$-th
highest valuation among the $N-1$ competitors facing a single bidder,
we obtain
\[
H(q)=\left\{ \begin{array}{ll}
G_{2}(q) & \text{if }q\geq\psi^{-1}(0)\\
G_{1}(q)+(N-1)\left[F(q+\psi(q))\right]^{N-2}\left[1-F(q)\right] & \text{if }\psi^{-1}(0)>q\geq a(0)\\
G_{1}(q) & \text{if }q<a(0),
\end{array}\right.
\]
where $(N-1)\left[F(q+\psi(q))\right]^{N-2}\left[1-F(q)\right]$ is
the probability of the event $Y_{(1)}>q$ and $\psi(q)+q\geq Y_{(2)}$
for $q\in\left[a(0),\psi^{-1}(0)\right)$.

Let $V(q,x)$ denote the expected payoff of a bidder of type $x$
who bids $\beta(q)$. When the bidder submits a truthful bid of $\beta(x)$,
then the probability that he gets a good is equal to the probability
that he pays his bid, and so his expected payoff is
\[
V(x,x)=\left[x-\beta(x)\right]H(x).
\]
If $x$ is the highest type, then he pays his bid to the first seller
and gets a refund from that seller for the price that he has to pay
to acquire the good from seller 2. The price depends on whether the
good is allocated, but his payoff does not. If $x$ is the second-highest
type, then he pays his bid only if the good is allocated to him and
gets zero otherwise.

If the bidder deviates and submits a bid different from $\beta(x)$,
then the probability of getting the good is no longer equal to $H(q)$
and the bidder's expected payoff includes additional terms, because
the outcome in the second auction is based on the true values of the
bidders. We show below that these deviations yield lower payoffs than
bidding truthfully. Further, those additional terms drop out when
we evaluate the derivative of $V(q,x)$ with respect to $q$ at $q=x$.
Taking that derivative yields the first-order condition
\begin{equation}
H^{\prime}(x)\left[x-\beta(x)\right]-H(x)\beta^{\prime}(x)=0.\label{eq:FOC}
\end{equation}
As in the standard mechanism design environment, we can find the equilibrium
bid function $\beta(\cdot)$ by solving the differential equation
in Expression \ref{eq:FOC}, with boundary condition $\beta(\underline{x})=0$.
To solve, rewrite Expression \ref{eq:FOC} as
\[
\frac{d}{dx}\left[H(x)\beta(x)\right]=xH^{\prime}(x).
\]
Integrating over an interval $[a,b]$ yields 
\begin{equation}
H(b)\beta(b)-H(a)\beta(a)=\int_{a}^{b}xH^{\prime}(x).\label{eq:integral}
\end{equation}

We then construct the solution piecewise by plugging in the values
of $H(\cdot)$ and $H^{\prime}(\cdot)$. For $x\in\left[\underline{x},a(0)\right)$,
\begin{equation}
\beta(x)=\frac{1}{G_{1}(x)}\int_{\underline{x}}^{x}sg_{1}(s).\label{eq:b low}
\end{equation}
For $x\in\left[a(0),\psi^{-1}(0)\right)$, 
\begin{equation}
\beta(x)=\frac{1}{H(x)}\left[G_{1}(a(0))\beta(a(0))+\int_{a(0)}^{x}sH^{\prime}(s)\right]\label{eq:b mid}
\end{equation}
where 
\[
H^{\prime}(s)=g_{1}(s)+(N-1)\left[\begin{array}{c}
(N-2)\left[F(s+\psi(s))\right]^{N-3}\left[1-F(s)\right](1+\psi^{\prime}(s))f(s+\psi(s))\\
-\left[F(s+\psi(s))\right]^{N-2}f(s)
\end{array}\right].
\]
Finally, for $x\in\left[\psi^{-1}(0),\bar{x}\right]$, 
\begin{equation}
\beta(x)=\frac{1}{G_{2}(x)}\left[G_{2}(\psi^{-1}(0))\beta(\psi^{-1}(0))+\int_{\psi^{-1}(0)}^{x}sg_{2}(s)\right].\label{eq:b high}
\end{equation}

We can now state our result that the modified pay-your-bid auction
implements the optimal mechanism.

\begin{theorem} \label{Theorem: modified first-price}If the distribution
of buyer values $F$ has increasing virtual valuations, then the bid
function specified in Expressions \ref{eq:b low}-\ref{eq:b high}
is an equilibrium of the modified pay-your-bid auction, and that equilibrium
yields the optimal expected revenue for the first seller. \end{theorem}

Because the bid function is invertible (it is strictly increasing),
the seller can implement the optimal reserve rule. Note that even
a buyer who has a valuation $x<a(0)$, and who therefore knows that
the mechanism will never assign him the first object, is willing to
participate. If his is the highest valuation, then he will obtain
the second good at a total cost equal to his bid $\beta(x)$ in the
first auction. (Recall that he pays $\beta(x)$ and then is refunded
the sale price $y_{(1)}$ in the second auction.) Since $x<a(0)$,
his bid is given by 
\[
\beta(x)=\frac{1}{G_{1}(x)}\int_{\underline{x}}^{x}y_{1}g_{1}(y_{1})=E[Y_{(1)}|Y_{(1)}\leq x],
\]
the expected sale price that he faces in the second auction, conditional
on having the highest valuation. Thus, he is willing to bid in the
first auction rather than wait for the second.

In the auction, the transfer from the second-highest bidder is always
positive. However, the realized transfer from the highest bidder may
be negative, in which case the first seller makes a payment to the
bidder. For example, suppose that the highest valuation $x_{(1)}$
is below $a(0)$ and that $x_{(2)}$ is close to $x_{(1)}$ (in particular,
$x_{(2)}>E[X_{(2)}|X_{(1)}=x_{(1)}]$). Then the item is not allocated,
and the highest bid, $\beta(x_{(1)})=E[X_{(2)}|X_{(1)}=x_{(1)}]$,
is less than the refund of the second auction's sale price, $x_{(2)}$:
$\beta(x_{(1)})-x_{(2)}<0.$

\subsection{Proof of Theorem \ref{Theorem: modified first-price}}

First, we show that the bid function $\beta$ specified in Expressions
\ref{eq:b low}-\ref{eq:b high} is an equilibrium. Suppose that buyer
$i$'s valuation is $x_{i}$ and that all other buyers are bidding
according to $\beta$. We want to show that submitting a bid of $\beta(x_{i})$
is a best response for buyer $i$. In particular, we want to show
that buyer $i$ cannot do better by submitting $\beta(q)$ for some
$q\in\left[\underline{x},\bar{x}\right]$. (Bidding above $\beta(\bar{x})$
or below $\beta(\underline{x})$ is dominated by bidding $\beta(\bar{x})$
or 0, respectively.)

For brevity, we will consider only the case where $x_{i}\geq\psi^{-1}(0)$
and $q>x_{i}$. The other cases are similar. If buyer $i$ submits
a bid of $\beta(x_{i})$, then his expected total (across both periods)
payoff is 
\begin{align}
 & \int_{\underline{x}}^{x_{i}}x_{i}g_{1}(y_{(1)})+\int_{x_{i}}^{\bar{x}}G_{2|y_{(1)}}(x_{i})x_{i}g_{1}(y_{(1)})-H(x_{i})\beta(x_{i})\nonumber \\
 & =H(x_{i})x_{i}-H(x_{i})\beta(x_{i}).\label{eq:true}
\end{align}
If he has the highest valuation, then he will pay the first seller
$\beta(x_{i})$, win the second auction at a price equal to the third-highest
valuation $x_{(3)}$, and get a refund of $x_{(3)}$ from the first
seller. If he has the second-highest valuation, then he will get the
first object (since $x_{i}>x_{H}$, so $x_{i}+\psi(x_{i})>x_{(3)}$)
and pay $\beta(x_{i})$.

Submitting a bid of $\beta(q)>\beta(x_{i})$ instead yields 
\begin{align}
 & \int_{\underline{x}}^{x_{i}}x_{i}g_{1}(y_{(1)})+\int_{x_{i}}^{q}G_{2|y_{(1)}}(x_{i})x_{i}g_{1}(y_{(1)})+\int_{q}^{\bar{x}}G_{2|y_{(1)}}(q)x_{i}g_{1}(y_{(1)})-H(q)\beta(q)\nonumber \\
 & \leq\int_{\underline{x}}^{q}x_{i}g_{1}(y_{(1)})+\int_{q}^{\bar{x}}G_{2|y_{(1)}}(q)x_{i}g_{1}(y_{(1)})-H(q)\beta(q)\nonumber \\
 & =H(q)x_{i}-H(q)\beta(q).\label{eq:exaggerate}
\end{align}
If he has the highest valuation, then again he will get the second
object for a total payment equal to his bid. If the highest valuation
among his competitors, $y_{(1)}$, is above $x_{i}$ but below $q$,
then the object will be allocated (because $y_{(1)}>x_{i}>\psi^{-1}(0)$)
and buyer $i$ will pay his bid $\beta(q)$, but he will win the second
auction only if the second-highest competitor's valuation, $y_{(2)}$,
is less than his true valuation $x_{i}$. If $q$ is the second-highest
bid, then buyer $i$ will get the first object and pay $\beta(q)$.

Subtracting Expression \ref{eq:exaggerate} from Expression \ref{eq:true},
we get that the difference in payoff between bidding $\beta(x_{i})$
and bidding $\beta(q)$ is greater than or equal to 
\[
\begin{array}{cl}
 & H(x_{i})x_{i}-H(q)x_{i}+H(q)\beta(q)-H(x_{i})\beta(x_{i})\\
= & H(x_{i})x_{i}-H(q)x_{i}+\int_{x_{i}}^{q}xH^{\prime}(x)\\
= & H(x_{i})x_{i}-H(q)x_{i}+H(q)q-H(x_{i})x_{i}-\int_{x_{i}}^{q}H(x)\\
= & H(q)(q-x_{i})-\int_{x_{i}}^{q}H(x)\\
= & \int_{x_{i}}^{q}\left(H(q)-H(x)\right)\\
\geq & 0,
\end{array}
\]
where the first equality uses Expression \ref{eq:integral} and the
second uses integration by parts. Thus, buyer $i$ cannot do better
by bidding above $\beta(x_{i})$. Similar arguments show that he cannot
do better by bidding below $\beta(x_{i})$ and that $\beta$ is a
best response for buyers with valuations below $\psi^{-1}(0)$ as
well.

\subsubsection{Revenue equivalence}

The proof that the modified first-price auction yields the optimal
expected revenue is similar to the one that Riley and Samuelson \cite{RileySamuelson}
use to show that auctions in a broad class generate the same expected
revenue in the standard mechanism design environment. As a preliminary,
we use the definitions of $G_{1}$ and $G_{2}$ to rewrite the probability
$H(q)$ that makes a payment in the first auction as 
\[
H(q)\equiv\begin{cases}
\lbrack F(q)]^{N-1}+(N-1)(1-F(q))[F(q)]^{N-2} & \mbox{if }q\geq\psi^{-1}(0)\\
\lbrack F(q)]^{N-1}+(N-1)(1-F(q))[F(q+y(q))]^{N-2} & \mbox{if }\psi^{-1}(0)>q\geq a(0)\\
\lbrack F(q)]^{N-1} & \mbox{if }q<a(0).
\end{cases}
\]

Let 
\[
P(x)\equiv H(x)\beta(x)
\]
denote the expected payment of a bidder with a valuation of $x$ (not
counting the rebate to the high bidder). We want to show that 
\[
\begin{array}{c}
N\cdot E\left[P(X)\right]\\
=\\
E\left[\psi(X_{(1)}\right]+E\left[\psi(X_{(2)}|X_{(2)}+\psi(X_{(2)})-X_{(3)}\geq0\right]\cdot\Pr\left(X_{(2)}+\psi(X_{(2)})-X_{(3)}\geq0\right).
\end{array}
\]

From Expression \ref{eq:FOC}, we know that equilibrium bids satisfy
the first-order condition
\[
xH^{\prime}(x)-P^{\prime}(x)=0
\]
for all $x\in\left[\underline{x},\bar{x}\right]$. Individual rationality
implies that $P(0)=0$, so we can integrate to get
\[
\begin{array}{cl}
P(x) & =\int_{0}^{x}x^{\prime}H^{\prime}(x^{\prime})\\
 & =xH(x)-\int_{0}^{x}H(x^{\prime}),
\end{array}
\]
where the second equality follows from integration by parts. Taking
the expectation over $x$ gives the \emph{ex ante} expected payment
from a bidder to the first seller:
\[
E\left[P(X)\right]=\int_{\underline{x}}^{\bar{x}}P(x)f(x)=\int_{0}^{\bar{x}}P(x)f(x).
\]
Substituting for $P(x)$ and integrating by parts, we obtain
\[
\begin{array}{cl}
E\left[P(X)\right] & =\int_{0}^{\bar{x}}\left[xH(x)-\int_{0}^{x}H(x^{\prime})dx^{\prime}\right]f(x)dx\\
 & =\int_{0}^{\bar{x}}\left[xf(x)+F(x)-1\right]H(x)dx.
\end{array}
\]
That is, $E\left[P(X)\right]$ equals 
\begin{equation}
\begin{array}{l}
=\int_{0}^{a(0)}\left[xf(x)+F(x)-1\right][F(x)]^{N-1}\\
+\int_{a(0)}^{\psi^{-1}(0)}\left[xf(x)+F(x)-1\right]\left([F(x)]^{N-1}+(N-1)(1-F(x))[F(x+\psi(x))]^{N-2}\right)\\
+\int_{\psi^{-1}(0)}^{\bar{x}}\left[xf(x)+F(x)-1\right]\left([F(x)]^{N-1}+(N-1)(1-F(x))[F(x)]^{N-2}\right).
\end{array}\label{eq:p-bar}
\end{equation}
The first line of Expression \ref{eq:p-bar} can be rewritten as
\[
\begin{array}{l}
\frac{1}{N}\int_{0}^{a(0)}\left[x-\frac{1-F(x)}{f(x)}\right]N[F(x)]^{N-1}f(x)\\
=\frac{1}{N}\int_{0}^{a(0)}\psi(x)f_{1}(x).
\end{array}
\]
(Recall that $f_{k}$ is the density of the $k$-th order statistic.)
Similarly, the third line of Expression \ref{eq:p-bar} equals
\[
\begin{array}{l}
\frac{1}{N}\int_{\psi^{-1}(0)}^{\bar{x}}\left[x-\frac{1-F(x)}{f(x)}\right]\left[N[F(x)]^{N-1}f(x)+N(N-1)(1-F(x))[F(x)]^{N-2}f(x)\right]\\
=\frac{1}{N}\int_{\psi^{-1}(0)}^{\bar{x}}\psi(x)f_{1}(x)+\frac{1}{N}\int_{\psi^{-1}(0)}^{\bar{x}}\psi(x)f_{2}(x).
\end{array}
\]
Finally, the second line of Expression \ref{eq:p-bar} equals
\[
\begin{array}{l}
\frac{1}{N}\int_{a(0)}^{\psi^{-1}(0)}\left[x-\frac{1-F(x)}{f(x)}\right]\left[N[F(x)]^{N-1}f(x)+N(N-1)(1-F(x))[F(x+\psi(x))]^{N-2}f(x)\right]\\
=\frac{1}{N}\int_{a(0)}^{\psi^{-1}(0)}\psi(x)f_{1}(x)+\frac{1}{N}\int_{x_{L}}^{x_{H}}\psi(x)f_{2}(x)\cdot\left(\frac{F(x+\psi(x))}{F(x)}\right)^{N-2}.
\end{array}
\]

Making those substitutions, we obtain that $N\cdot E\left[P(X)\right]$
is equal to
\[
\int_{0}^{\bar{x}}\psi(x)f_{1}(x)+\int_{a(0)}^{\bar{x}}\psi(x)f_{2}(x)\cdot\min\left\{ 1,\left(\frac{F(x+\psi(x))}{F(x)}\right)^{N-2}\right\} .
\]
Recall that the object is allocated when $x_{(3)}\leq x_{(2)}+\psi(x_{(2)})$.
As desired, then, the sum of expected payments is exactly the expectation
of the highest virtual valuation $\psi(X_{(1)})$, plus the expectation
of the second-highest virtual valuation $\psi(X_{(2)})$ weighted
by the probability that the object will be assigned given the value
of $X_{(2)}$.

\section{Second-Price Auction with Reserve Price\label{sec:Thresholds-in-Second-Price}}

To characterize the equilibrium, we introduce some notation. For $k\in\{0,1,2\}$,
define $p_{k}(\hat{x},\hat{\hat{x}})$ as the probability that a buyer
has exactly $k$ rivals with types between $\hat{x}$ and $\hat{\hat{x}}$
and no rivals with a higher type: 
\[
p_{0}=\left[F(\hat{x})\right]^{N-1}\left(=G_{1}(\hat{x})\right)=(\hat{x})^{2},
\]
\[
p_{1}=(N-1)\left[F(\hat{\hat{x}})-F(\hat{x})\right]\left[F(\hat{x})\right]^{N-2}=2(\hat{\hat{x}}-\hat{x})\hat{x},
\]
\[
p_{2}=\frac{(N-1)(N-2)}{2}\left[F(\hat{\hat{x}})-F(\hat{x}\right]^{2}\left[F(\hat{x})\right]^{N-3}=(\hat{\hat{x}}-\hat{x})^{2}.
\]
Define $D_{k}$ as the expected value of the highest rival type in
the second auction, conditional on $k$ and conditional on one of
the $k$ rivals winning the first auction if $k>0$: 
\[
D_{0}=E\left[y_{1}|y_{1}<\hat{x}\right]=\frac{2}{3}\hat{x},
\]
\[
D_{1}=E\left[y_{2}|y_{1}\in\left[\hat{x},\hat{\hat{x}}\right],y_{2}<\hat{x}\right]=\frac{1}{2}\hat{x},
\]
\[
D_{2}=\frac{1}{2}E\left[y_{1}|y_{1},y_{2}\in\left[\hat{x},\hat{\hat{x}}\right]\right]+\frac{1}{2}E\left[y_{2}|y_{1},y_{2}\in\left[\hat{x},\hat{\hat{x}}\right]\right]=\frac{1}{2}(\hat{\hat{x}}+\hat{x}).
\]
Finally, for $x\in\lbrack\hat{x},\hat{\hat{x}}]$, let 
\[
L(x)\equiv\int_{\hat{\hat{x}}}^{\bar{x}}\left[\int_{\underline{x}}^{x}[x-y_{(2)}]g_{2|y_{(1)}}(y_{(2)})\right]g_{1}(y_{(1)})
\]
denote the expected payoff in the second auction conditional on the
winner of the first auction having a type above $\hat{\hat{x}}$,
times the probability of that event. The dependence of $p_{k}$, $D_{k}$,
and $L(x)$ on $\hat{x}$ and $\hat{\hat{x}}$ is suppressed for readability.

The cutoff values $\hat{x}$ and $\hat{\hat{x}}$ are characterized
by two indifference conditions. A buyer of type $\hat{\hat{x}}$ is
indifferent between bidding $r$ (and tying with other types in $[\hat{x},\hat{\hat{x}}]$)
and bidding just above $r$; a buyer of type $\hat{x}$ is indifferent
between bidding $r$ and not bidding. That is, 
\begin{align*}
 & p_{0}(\hat{\hat{x}}-r)+p_{1}\left[\frac{1}{2}(\hat{\hat{x}}-r)+\frac{1}{2}(\hat{\hat{x}}-D_{1}\right]+p_{2}\left[\frac{1}{3}(\hat{\hat{x}}-r)+\frac{2}{3}(\hat{\hat{x}}-D_{2})\right]+L(\hat{\hat{x}})\\
 & =p_{0}(\hat{\hat{x}}-r)+p_{1}(\hat{\hat{x}}-r)+p_{2}(\hat{\hat{x}}-r)+L(\hat{\hat{x}})
\end{align*}
and 
\begin{align*}
 & p_{0}(\hat{x}-r)+p_{1}\left[\frac{1}{2}(\hat{x}-r)+\frac{1}{2}(\hat{x}-D_{1}\right]+p_{2}\frac{1}{3}(\hat{x}-r)+L(\hat{\hat{x}})\\
 & =p_{0}(\hat{x}-D_{0})+p_{1}(\hat{x}-D_{1})+p_{2}\cdot0+L(\hat{\hat{x}}).
\end{align*}
Taking differences, $\hat{\hat{x}}$ solves 
\[
p_{1}\frac{1}{2}(D_{1}-r)+p_{2}\frac{2}{3}(D_{2}-r)=0
\]
and $\hat{x}$ solves 
\[
p_{0}(D_{0}-r)+p_{1}\frac{1}{2}(D_{1}-r)+p_{2}\frac{1}{3}(\hat{x}-r)=0.
\]

Plugging in the values of $p_{k}$ and $D_{k}$ gives
\begin{equation}
\hat{x}\left(\frac{1}{2}\hat{x}-r\right)+(\hat{\hat{x}}-\hat{x})\frac{2}{3}\left(\frac{1}{2}(\hat{\hat{x}}+\hat{x})-r\right)=0\label{eq:x*, x** first}
\end{equation}
and 
\begin{equation}
(\hat{x})^{2}\left(\frac{2}{3}\hat{x}-r\right)+(\hat{\hat{x}}-\hat{x})\hat{x}\left(\frac{1}{2}\hat{x}-r\right)+\frac{1}{3}(\hat{\hat{x}}-\hat{x})^{2}(\hat{x}-r)=0.\label{eq:x*, x** second}
\end{equation}
The solutions to Expressions \ref{eq:x*, x** first} and \ref{eq:x*, x** second}
are $\hat{x}=(1+1/\sqrt{3})r$ and $\hat{\hat{x}}=(1+2/\sqrt{3})r$.

The revenue of the first seller is
\begin{align*}
R_{1}(r_{1}) & =\left[F_{1}(\hat{\hat{x}})-F_{1}(\hat{x})\right]r_{1}+\int\limits _{\hat{\hat{x}}}^{1}\left[F_{2|x_{(1)}}(\hat{\hat{x}})r_{1}+\int\limits _{\hat{\hat{x}}}^{x_{(1)}}\beta(x_{(2)})f_{2|x_{(1)}}(x_{(2)})\right]f_{1}(x_{(1)})\\
 & =\left[(\hat{\hat{x}})^{3}-(\hat{x})^{3}\right]r_{1}+\int\limits _{\hat{\hat{x}}}^{1}\left[\frac{(\hat{\hat{x}})^{2}}{(x_{(1)})^{2}}r_{1}+\int\limits _{\hat{\hat{x}}}^{x_{(1)}}\frac{x_{(2)}}{2}\frac{2x_{(2)}}{(x_{(1)})^{2}}\right]3(x_{(1)})^{2}\\
 & =\frac{1}{4}-(\hat{\hat{x}})^{3}+\frac{3}{4}(\hat{\hat{x}})^{4}+r_{1}\left[3(\hat{\hat{x}})^{2}-2(\hat{\hat{x}})^{3}-(\hat{x})^{3}\right],
\end{align*}
where the last line follows from a lot of tedious calculations. Plugging
in $\hat{x}=(1+1/\sqrt{3})r$ and $\hat{\hat{x}}=(1+2/\sqrt{3})r$
and performing more tedious calculation gives
\[
R_{1}(r_{1})=\frac{1}{4}+r_{1}^{3}\left[\frac{6\sqrt{3}+10}{3\sqrt{3}}\right]-r_{1}^{4}\left[\frac{47\sqrt{3}+80}{12\sqrt{3}}\right].
\]
We can now determine the optimal reserve price. Differentiating with
respect to $r_{1}$ yields the first-order condition
\[
3r_{1}^{2}\left[\frac{6\sqrt{3}+10}{3\sqrt{3}}\right]-4r_{1}^{3}\left[\frac{47\sqrt{3}+80}{12\sqrt{3}}\right]=0.
\]
Solving for the optimal reserve price $r_{1}^{\ast}$ yields
\[
r_{1}^{\ast}=\frac{3\left[6\sqrt{3}+10\right]}{\left[47\sqrt{3}+80\right]}\approx0.379.
\]
The corresponding values of $\hat{x}$ and $\hat{\hat{x}}$ are $\hat{x}=(1+1/\sqrt{3})r_{1}^{\ast}\approx0.60$
and $\hat{\hat{x}}=(1+2/\sqrt{3})r_{1}^{\ast}\approx0.82$.

Substituting the optimal reserve into the revenue function yields
the maximal revenue, which is
\begin{align*}
R_{1}(r_{1}^{\ast}) & =\frac{1}{4}+(r_{1}^{\ast})^{3}\left[\frac{6\sqrt{3}+10}{3\sqrt{3}}\right]-(r_{1}^{\ast})^{4}\left[\frac{47\sqrt{3}+80}{12\sqrt{3}}\right]\\
 & =\frac{1}{4}+\frac{27}{256}\frac{\left(\frac{6\sqrt{3}+10}{3\sqrt{3}}\right)^{4}}{\left(\frac{47\sqrt{3}+80}{12\sqrt{3}}\right)^{3}}\approx0.303.
\end{align*}

We can also compute the expected revenue for the second seller when
the first seller sets the optimal reserve price. She gets the second-highest
valuation $x_{(2)}$ if the first seller does not allocate and the
third-highest valuation $x_{(3)}$ otherwise -- except if all three
valuations are between $\hat{\hat{x}}$ and $\hat{\hat{x}}$ and the
first seller randomly allocates to the buyer with valuation $x_{(3)}$,
in which case the second seller gets $x_{(2)}$ instead of $x_{(3)}$:
\begin{align*}
R_{2}(r_{1}^{\ast}) & =\int_{0}^{\hat{x}}E\left[X_{(2)}|X_{(1)}=x_{(1)}\right]f_{1}(x_{(1)})+\int_{\hat{x}}^{\overline{x}}E\left[X_{(3)}|X_{(1)}=x_{(1)}\right]f_{1}(x_{(1)})\\
 & +\frac{1}{3}E\left[X_{(2)}-X_{(3)}|X_{(1)},X_{(2)},X_{(3)}\in\left[\hat{x},\hat{\hat{x}}\right]\right]\Pr\left[X_{(1)},X_{(2)},X_{(3)}\in\left[\hat{x},\hat{\hat{x}}\right]\right]\\
 & =\int_{0}^{\hat{x}}\frac{2}{3}x_{(1)}3(x_{(1)})^{2}+\int_{\hat{x}}^{\overline{x}}\frac{1}{3}x_{(1)}3(x_{(1)})^{2}\\
 & +\frac{1}{3}\left[\left(\frac{1}{2}\hat{x}+\frac{1}{2}\hat{\hat{x}}\right)-\left(\frac{3}{4}\hat{x}+\frac{1}{4}\hat{\hat{x}}\right)\right]\left(\hat{\hat{x}}-\hat{x}\right)^{3}\\
 & =\frac{1}{4}+\frac{1}{4}(\hat{x})^{4}-\frac{1}{12}\left(\hat{\hat{x}}-\hat{x}\right)^{4}\approx0.282.
\end{align*}


\begin{thebibliography}{10}
\bibitem{Ashenfelter}Ashenfelter, O. (1989). ``How Auctions Work
for Wine and Art,\textquotedblright \ \emph{Journal of Economic Perspectives},
3: 23-36.

\bibitem{AshenfelterGenesove}Ashenfelter, O. and D. Genesove (1992).
``Testing for Price Anomalies in Real Estate Auctions,\textquotedblright \ \emph{American
Economic Review Papers and Proceedings}, 82: 501-5.

\bibitem{AshenfelterGraddy}Ashenfelter, O. and K. Graddy (2003).
``Auctions and the Price of Art,\textquotedblright \ \emph{Journal
of Economic Literature}, 41: 763-87.

\bibitem{BackusLewis}Backus, M. and G. Lewis (2019), ``Dynamic Demand
Estimation in Auction Markets,\textquotedblright \ working paper.

\bibitem{BeggsGraddy}Beggs, A. and K. Graddy (1997). ``Declining
Values and the Afternoon Effect: Evidence from Art Auctions,\textquotedblright \ \emph{RAND
Journal of Economics}, 28: 544-65.

\bibitem{BBM}Bergemann, D., B. Brooks, and S. Morris (Forthcoming).
``Countering the Winner\textquoteright s Curse: Optimal Auction Design
in a Common Value Model,'' \emph{Theoretical Economics}.

\bibitem{BlackdeMeza}Black, J. and D. de Meza (1992). ``Systematic
Price Differences Between Successive Auctions are No Anomaly,\textquotedblright \ \emph{Journal
of Economics \& Management Strategy}, 1: 607-28.

\bibitem{Brugues}Brugués, J. (2020). ``The Effects of Public Procurement
on Medicine Supply,'' working paper.

\bibitem{BudishZeithammer}Budish, E. and R. Zeithammer (2016). ``An
Efficiency Ranking of Markets Aggregated from Single-Object Auctions,\textquotedblright \ working
paper.

\bibitem{BurguetSakovics}Burguet, R. and J. Sákovics (1999). ``Imperfect
Competition in Auction Designs,\textquotedblright \ \emph{International
Economic Review}, 40: 231-247.

\bibitem{CalzolariPavan}Calzolari, G. and A. Pavan (2006). ``Monopoly
with Resale,\textquotedblright \ \emph{RAND Journal of Economics},
37: 362-375.

\bibitem{CarrollSegal}Carroll, G. and I. Segal (2019). ``Robustly
Optimal Auctions with Unknown Resale Opportunities,\textquotedblright \ \emph{Review
of Economic Studies}, 86: 1527-1555.

\bibitem{Dworczak}Dworczak, P. (2020). ``Mechanism Design with Aftermarkets:
Cutoff Mechanisms,\textquotedblright \ working paper.

\bibitem{FigueroaSkreta}Figueroa, N. and V. Skreta (2009). ``A Note
on Optimal Allocation Mechanisms,\textquotedblright \ \emph{Economic
Letters}, 102: 169-173.

\bibitem{JehielMoldovanu}Jehiel, P. and B. Moldovanu (2003). ``Auctions
with Downstream Interaction Among Buyers,\textquotedblright \ \emph{RAND
Journal of Economics}, 31: 768-91.

\bibitem{JehielMoldovanuStacchetti1996}Jehiel, P., B. Moldovanu,
and E. Stacchetti (1996). ``How (Not) to Sell Nuclear Weapons,\textquotedblright \ \emph{American
Economic Review, }86: 814-829.

\bibitem{JehielMoldovanuStacchetti1999}Jehiel, P., B. Moldovanu,
and E. Stacchetti (1999). ``Multidimensional Mechanism Design for
Auctions with Externalities,\textquotedblright \ \emph{Journal of
Economic Theory}, 85: 258-293.

\bibitem{KirkegaardOvergaard}Kirkegaard, R. and P.B. Overgaard (2008).
``Buy-Out Prices in Auctions: Seller Competition and Multi-Unit Demands,\textquotedblright \ \emph{RAND
Journal of Economics}, 39: 770-789.

\bibitem{Krishna}Krishna, V. (2010). \textbf{Auction Theory}, Second
Edition. San Diego: Academic Press.

\bibitem{LoertscherMarx}Loertscher, S. and L. Marx (2020). ``Asymptotically
Optimal Prior-Free Clock Auctions,'' \emph{Journal of Economic Theory},
187: 105030.

\bibitem{McAfee}McAfee, P. (1993). ``Mechanism Design by Competing
Sellers,\textquotedblright \ \emph{Econometrica,} 61: 1281-1312.

\bibitem{MilgromWeber}Milgrom, P. and R. Weber (2000). ``A Theory
of Auctions and Competitive Bidding II,\textquotedblright \ in \textbf{The
Economic Theory of Auctions}, Paul Klemperer (ed.). Cheltenham, UK:
Edward Elgar Publishing.

\bibitem{Myerson}Myerson, R. (1981). ``Optimal Auction Design,\textquotedblright \ \emph{Mathematics
of Operations Research}, 6: 58-73.

\bibitem{Pai}Pai, M. (2009). ``Competing Auctioneers,'' working
paper.

\bibitem{Peters}Peters, M. (2001). ``Surplus Extraction and Competition,\textquotedblright \ \emph{Review
of Economic Studies}, 68: 613-631.

\bibitem{Peters2}Peters, M. (2010). ``Competing Mechanisms,\textquotedblright \ working
paper.

\bibitem{PetersSeverinov}Peters, M. and S. Severinov (1997). ``Competition
Among Sellers who offer Auctions instead of Prices,\textquotedblright \ \emph{Journal
of Economic Theory}, 75: 141-179.

\bibitem{RileySamuelson}Riley, J. and W. Samuelson (1981). ``Optimal
Auctions,\textquotedblright \ \emph{American Economic Review}, 71:
381-92.

\bibitem{Said}Said, M. (2011). ``Sequential Auctions with Randomly
Arriving Buyers,\textquotedblright \ \emph{Games and Economic Behavior},
73: 238-243.

\bibitem{Virag}Virág, G. (2016). ``Auctions with Resale: Reserve
Prices and Revenues,'' \emph{Games and Economic Behavior}, 99: 239-249.

\bibitem{Zeithammer}Zeithammer, R. (2006). ``Forward-Looking Bidding
in Online Auctions,\textquotedblright \ \emph{Journal of Marketing
Research}, 43: 462-476.
\end{thebibliography}
\end{document}